\documentclass[dvi,11pt,a4paper]{article}
\usepackage{jcappub}
\usepackage{float}
\usepackage{multirow}
\usepackage{graphicx}
\usepackage{slashbox}
\usepackage{bm}

\newcommand{\squeezeup}{\vspace{-5.0mm}}

\title{Confronting Recent Results from Selected Direct and Indirect Dark Matter Searches and the Higgs Boson with Supersymmetric Models with Non-universal Gaugino Masses}

\author[a]{A. Spies,}
\author[a]{G. Anton}

\affiliation[a]{Erlangen Center for Astroparticle Physics, Department of Physics, Friedrich-Alexander-University of Erlangen-Nuremberg}

\emailAdd{andreas.spies@physik.uni-erlangen.de}
\emailAdd{gisela.anton@physik.uni-erlangen.de}

\abstract{In this paper we study a class of supersymmetric models with non-universal gaugino masses that arise from a mixture of SU(5) singlet and non-singlet representations, i.e. a combination of \textbf{1}, \textbf{24}, \textbf{75} and \textbf{200}. Based on these models we calculate the expected dark matter signatures within the linear combination \textbf{1} $\oplus$ \textbf{24} $\oplus$ \textbf{75} $\oplus$ \textbf{200}. We confront the model predictions with the detected boson as well as current experimental limits from selected indirect and direct dark matter search experiments ANTARES respective IceCube and XENON. We comment on the detection/exclusion capability of the future XENON 1t project. For the investigated parameter span we could not find a SU(5) singlet model that fulfils the Higgs mass and the relic density constraint. In contrary, allowing a mixture of \textbf{1} $\oplus$ \textbf{24} $\oplus$ \textbf{75} $\oplus$ \textbf{200} enables a number of models fulfilling these constraints.}

\keywords{Supersymmetry, SU(5), non-universal, Gaugino, Higgs Boson, direct detection, indirect detection, XENON, Antares, IceCube}

\begin{document}
\maketitle
\flushbottom

\section{Introduction}
Supersymmetric extensions of the standard model (SM) of particle physics provide an elegant way to solve some major problems of the SM, e.g. by explaining the huge gap between the weak scale and the Planck scale. Moreover, if R-parity is conserved, supersymmetry (SUSY) can provide a dark matter candidate in a natural way. In most SUSY scenarios, the lightest neutralino $\chi_{1}^{0}$ is the lightest supersymmetric particle (LSP). Being a massive, electrical neutral, stable and weakly interacting particle makes it an excellent candidate to explain dark matter from the particle physics point of view. Throughout this paper we refer to $\chi_{1}^{0}$ as the neutralino $\chi$. \\
\indent Several observations constrain the parameter space of SUSY models. Such constraints are the correct parameter of electro-weak symmetry breaking or the relic dark matter density determined from WMAP data. Recently, the LHC experiments reported the discovery of a particle of 125 GeV/$c^{2}$ mass which very probably is a Higgs boson \cite{bib:CMSHiggs}, \cite{bib:ATLASHiggs}. This Higgs particle will further constrain the SUSY parameter space.\\
\indent In this paper we study a class of supersymmetric models introduced in \cite{bib:JEYounkinSPMartin} where a mixture between the singlet representation \textbf{1} of the gauge group SU(5) and its adjoint representation \textbf{24} was considered. It was shown, that these models provide a solution to the ''little hierarchy problem'' of supersymmetry and a neutralino dark matter candidate.\\
\indent While in \cite{bib:JEYounkinSPMartin} only a rather limited parameter range was consistent with the constraints, especially with constraints coming from the mass of the Higgs boson, we try to find more extended parameter regions that simultaneously incorporate a Higgs boson with the correct mass (within theoretical and experimental uncertainties) and an explanation for dark matter. We adapt the parameterization of \cite{bib:JEYounkinSPMartin} and extend it to the more general case of \textbf{1} $\oplus$ \textbf{24} $\oplus$ \textbf{75} $\oplus$ \textbf{200}.\\
\indent We investigate the phenomenological consequences with respect to dark matter search experiments. Two classes of experiments are considered: direct detection experiments which search for signatures produced by scattering a dark matter particle off atomic nuclei while indirect detection experiments search for self annihilation products of dark matter particles. We refer to the results published by the direct detection experiment XENON and to the indirect detection experiments ANTARES and IceCube.\\
\indent This paper is organized as follows. In Section 2 we introduce the parameterization used in this analysis. We comment on the recent experimental results from the LHC concerning a Higgs particle and its implications for dark matter searches in Section 3. Section 4 is dedicated to the phenomenological implications of the parameterization introduced in Section 2 with respect to the relic density resulting from such models. In Section 5 we comment on indirect detection observables, i.e. the muon neutrino flux and related muon flux from neutralino annihilations in the Sun with respect to indirect detection experiments ANTARES and IceCube with its low energy extension DeepCore (ICDC). The direct detection experiment XENON is part of our discussion in Section 6. The excluded regions of the parameter space from the combined results of IceCube and XENON 100 is presented in Section 7. In Section 8 we present a preliminary study of models with non-universal gaugino masses. We deviate from gaugino mass ratios given by SU(5) group theoretical factors, allowing for variable gaugino mass ratios. Our conclusions are presented in Section 9.

\section{Parameterization of SU(5) Singlet and Non-singlet Combination}
Non-universalities in the gaugino sector arise from a chiral function $f_{ab}(\phi^{I})$ in the gauge kinetic part of the Lagrangian $\mathcal{L}_{\mathrm{gauge kin}}$ \cite{bib:CremmerEtAl}. $\phi^{I}$ is a chiral superfield and $\mathcal{L}_{\mathrm{gauge kin}}$ is given by \cite{bib:HuituEtAl}

\begin{equation}
  \mathcal{L}_{\mathrm{gauge kin}} = -\frac{1}{4} Re f_{ab}(\varphi^{I}) F_{\mu\nu}^{a} F^{b\mu\nu} + F_{a^{\prime}b^{\prime}}^{I} \frac{ \partial f_{ab}(\varphi^{I}) }{ \partial \varphi^{I}_{a^{\prime}b^{\prime}} } \lambda^{a}\lambda^{b} + H.c. + ...
\end{equation}

\noindent where $a$, $b$ are indices of the gauge generators, $\phi^{I}$'s denote the chiral superfields and $\lambda^{a}$ is the SU(5) gaugino field. $\varphi^{I}$ is the scalar component of $\phi^{I}$ whereas $F^{I}$ is its auxiliary F-component. $f_{ab}$ transforms as the symmetric product of two adjoint representations

\begin{equation}\label{ClebschGordanSeries}
  (\textbf{24} \otimes \textbf{24})_{\mathrm{symmetric}} = \textbf{1} \oplus \textbf{24} \oplus \textbf{75} \oplus \textbf{200}
\end{equation}

\noindent and is given by \cite{bib:HuituEtAl}

\begin{equation}
  f_{ab}(\phi^{I}) = f_{0} (\phi^{\mathrm{singlet}}) \delta_{ab} + \zeta_{\mathrm{Mult}} (\phi^{\mathrm{singlet}}) \frac{ \phi^{\mathrm{Mult}}_{ab} }{M_{\mathrm{Planck}}} + \mathcal{O} (( \frac{\phi^{\mathrm{Mult}}_{ab}}{M_{\mathrm{Planck}}} )^{2})
\end{equation}

\noindent In the above equations Einsteins sum convention was used for indices appearing twice. $f_{0}$ and $\zeta_{\mathrm{Mult}}$ are functions of gauge singlets $\phi^{\mathrm{singlet}}$. The index ''Mult'' labels possible multiplets of Eq. \ref{ClebschGordanSeries}, that are allowed as a linear term of $\phi^{\mathrm{Mult}}$ in $f_{ab}(\phi^{I})$. Supersymmetry is broken by the F-components $F^{I}$ of the chiral superfields $\phi^{I}$ when they acquire non-zero vevs and thus gaugino masses are generated. For the case of a non-singlet these gaugino masses ($M_{1}$, $M_{2}$, $M_{3}$) are unequal but related to each other \cite{bib:EllisEtAl}. Their relative magnitude at the scale of grand unification is given by group theoretical factors according to \cite{bib:AmundsonEtAl} 

\begin{equation}
  \left\langle F_{\phi} \right\rangle_{ab} = c_{a}\delta_{ab}
\end{equation}

\noindent with the coefficients $c_{a}$ listed in Table \ref{tbl:massRatioSU5Gut}

\begin{table}[h]
 \centering
 \begin{tabular}{|| c || c | c | c ||}
  \hline
  Representation & $M_{3}(M_{GUT})$ & $M_{2}(M_{GUT})$ & $M_{1}(M_{GUT})$ \\
  \hline
  \textbf{1} & 1 & 1 & 1 \\
  \textbf{24} & -2 & 3 & 1 \\
  \textbf{75} & -1 & -3 & 5 \\
  \textbf{200} & 1 & 2 & 10 \\
  \hline
 \end{tabular}
 \caption{SU(5) mass ratios (coefficients $c_{a}$) at the GUT scale for \textbf{1}, \textbf{24}, \textbf{75} and \textbf{200} representation of SU(5)}
 \label{tbl:massRatioSU5Gut}
\end{table}

\noindent A mixture of singlet and non-singlet representations can be written in form of three non-universality equations for $M_{1}$, $M_{2}$ and $M_{3}$

\begin{align}\label{eq:spmparaGeneral}
 M_{1} & =  m_{1/2} \left( \mathrm{cos}\left( \theta_{1} \right) + \sum_{i} a_{i} \ \mathrm{sin}\left( \theta_{i} \right) \right) \nonumber \\
 M_{2} & =  m_{1/2} \left( \mathrm{cos}\left( \theta_{1} \right) + \sum_{i} b_{i} \ \mathrm{sin}\left( \theta_{i} \right) \right) \nonumber \\
 M_{3} & =  m_{1/2} \left( \mathrm{cos}\left( \theta_{1} \right) + \sum_{i} c_{i} \ \mathrm{sin}\left( \theta_{i} \right) \right) 
\end{align}

\noindent where $i=24,75,200$ labels the possible multiplets, $(a_{24},a_{75},a_{200})=(1,5,10)$, $(b_{24},b_{75},b_{200})=(3,-3,2)$ and $(c_{24},c_{75},c_{200})=(-2,-1,1)$ . $\theta_{1}$ reflects the contribution of the singlet. $\theta_{i}$ reflects the contribution of the corresponding multiplet to the non-universality of the model. If $\theta_{1} = 0$ and all $\theta_{i} = 0$ we obtain the cMSSM scenario (often referred to as mSUGRA) where $M_{1} = M_{2} = M_{3} = m_{1/2}$. For $\theta_{1} = \pi/2$ and all $\theta_{i} = \pi/2$ we have a pure $SU(5)$ non-singlet contribution reflecting the given mass ratios of Table \ref{tbl:massRatioSU5Gut}.\\
\indent The above parameterization was adapted from \cite{bib:JEYounkinSPMartin}. There, only \textbf{1} $\oplus$ \textbf{24} was considered. Instead of coefficients $a_{i}$, $b_{i}$ and $c_{i}$ only coefficients $a_{24}$, $b_{24}$, and $c_{24}$ with $(a_{24},b_{24},c_{24})=(1,3,-2)$ occur in Equation \ref{eq:spmparaGeneral}. Moreover, the analysis of \cite{bib:JEYounkinSPMartin} imposed $\theta_{1} = \theta_{24}$. We will refer to that model briefly in the next Section. In total we obtain a 9 dimensional parameter space

\begin{align}\label{eq:spmParamSpace}
 m_{0} & = \textnormal{unified mass of scalars} \nonumber \\
 m_{1/2} & = \textnormal{gaugino mass parameter} \nonumber \\
 A_{0} & = \textnormal{unified trilinear couplings} \nonumber \\
 \mathrm{\mathrm{tan}}\beta & = \textnormal{ratio of Higgs vacuum expectation values} \nonumber \\
 \mathrm{sign}(\mu) & = \textnormal{sign of Higgs mass parameter } \mu \ +1 \textnormal{or} \ -1 \nonumber \\
 \theta_{1} & = \textnormal{contribution of the singlet} \nonumber \\
 \theta_{24} & = \textnormal{contribution of the 24-plet} \nonumber \\
 \theta_{75} & = \textnormal{contribution of the 75-plet} \nonumber \\
 \theta_{200} & = \textnormal{contribution of the 200-plet}
\end{align} 

\noindent We restricted the 5 parameters $m_{0}$, $m_{1/2}$, $A_{0}$, $\mathrm{tan}\beta$, $\mathrm{sign}(\mu)$ to the region which was evaluated in \cite{bib:JEYounkinSPMartin} in order to be able to investigate how far a generalised mixing ($\theta_{i} \neq 0$) changes the results. Accordingly our simulations use the following parameter range:

\begin{align}
 & 0 < m_{0} < 5 \ \textnormal{TeV} \nonumber \\
 & m_{1/2} = 600 \mathrm{GeV} \nonumber \\
 & A_{0} = -m_{1/2} \nonumber \\
 & \mathrm{tan}\beta = 10 \ \textnormal{resp.} \ 45 \nonumber \\
 & \mathrm{sign}(\mu) = +1 \nonumber \\
 & -0.25 < \theta_{1}/\pi < 0.75 \nonumber \\
 & -0.25 < \theta_{24}/\pi < 0.75 \nonumber \\
 & -0.25 < \theta_{75}/\pi < 0.75 \nonumber \\
 & -0.25 < \theta_{200}/\pi < 0.75
\end{align}

\noindent When we use a different set of parameters it is explicitely mentioned in the text. Our results for the linear combination \textbf{1} $\oplus$ \textbf{24} are in good agreement with those given in \cite{bib:JEYounkinSPMartin}. We found the combinations \textbf{1} $\oplus$ \textbf{75} and \textbf{1} $\oplus$ \textbf{200} provide less models fulfilling the constraints of the following sections compared to the parameterization of Younkin and Martin \cite{bib:JEYounkinSPMartin}. In principle any of the representations appearing in the symmetric product (Eq. \ref{ClebschGordanSeries}) must be treated equal and non of them should be preferred.\\
\indent To calculate the supersymmetric particle spectrum we used the public code SuSpect \cite{bib:DjouadiKneurMoultaka}. DarkSUSY \cite{bib:GondoloEdsjoBaltz} was employed for simulating dark matter observables. As such observables we investigated the muon neutrino flux $\phi_{\nu_{\mu}}$ and the resulting muon flux $\phi_{\mu}$ for indirect dark matter detection. As signal of direct detection we calculated the spin independent WIMP nucleon cross-section, $\sigma_{\mathrm{SI}}^{\mathrm{nucleon}}$.
\section{Higgs candidate from LHC results}
\begin{figure}[!ht]
  \begin{minipage}[t]{.48\textwidth}
    \begin{center}
      \includegraphics[width=8.0cm]{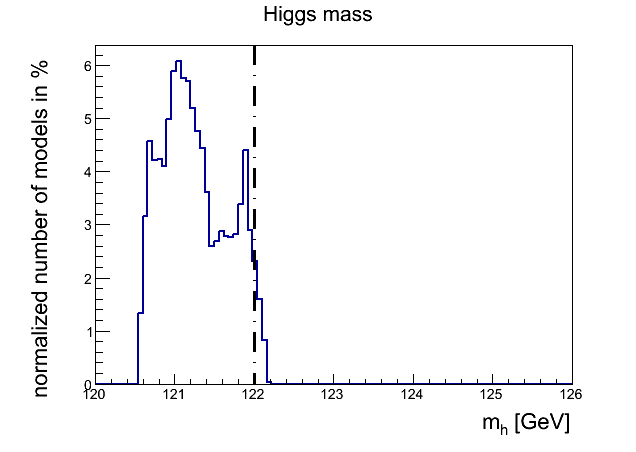}
    \end{center}
  \end{minipage}
  \hfill
  \begin{minipage}[t]{.48\textwidth}
    \begin{center}
      \includegraphics[width=8.0cm]{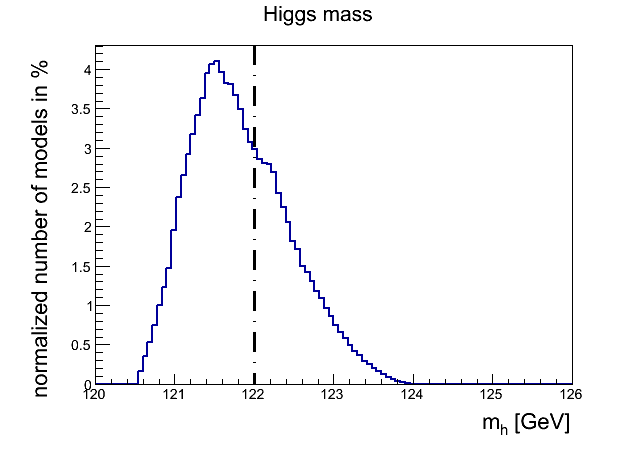}
    \end{center}
  \end{minipage}
  \caption[Predicted mass of the Higgs boson for \textbf{1} $\oplus$ \textbf{24} and \textbf{1} $\oplus$ \textbf{24} $\oplus$ \textbf{75} $\oplus$ \textbf{200}]{Number of models with a calculated Higgs mass $m_{h}$ is plotted on the y-axis for representations \textbf{1} $\oplus$ \textbf{24} (left) and \textbf{1} $\oplus$ \textbf{24} $\oplus$ \textbf{75} $\oplus$ \textbf{200} (right) of SU(5) for $m_{0} = 4$ TeV, $m_{1/2} = 600$ GeV, $A_{0} = -m_{1/2}$ and $\mathrm{tan}\beta=$10, the corresponding Higgs mass is plotted on the x-axis; black dash-dotted line: lower limit of the theoretical $\pm$ 3 GeV uncertainty with respect to the measurements of \cite{bib:CMSHiggs} and \cite{bib:ATLASHiggs} of $m_{h} \sim 125$ GeV. The number of models is normalized to the total number of simulated models.}
  \label{fig:su5Rep24And300FixM12TanBeta10HiggsMassDist}
\end{figure}

Last year, LHC experiments reported the discovery of a new boson with mass of 125 GeV/$c^{2}$ which might turn out to be the Higgs boson \cite{bib:CMSHiggs}, \cite{bib:ATLASHiggs}. In order to take theoretical uncertainties of the calculated mass of the Higgs boson into account we allow an uncertainty of $\pm$ 3 GeV/$c^{2}$ \cite{bib:BAllanachADjouadiEtAl} on the mass of the Higgs boson. The implications of such a Higgs boson on the kind of models investigated in this paper is discussed in this Section.\\
\indent Younkin and Martin indicated that only a few models survive the experimental and theoretical constraints from the Higgs boson. In their simulations they kept the gluino mass parameter $M_{3}$ and subsequently the bino mass parameter $M_{1}$ fixed. Here, we use a slightly different parameterization for \textbf{1} $\oplus$ \textbf{24}. Instead of fixing $M_{3}$ and $M_{1}$, we fix the overall gaugino mass scale $m_{1/2} = 600$ GeV (see Section 2). In contrast to Younkin et al, we simulated the model predictions with respect to the Higgs mass for independently varying the singlet mixing angle, $\theta_{1}$, and the 24-plet mixing angle $\theta_{24}$. The predicted distribution of the Higgs mass is shown on the left hand side of Figure \ref{fig:su5Rep24And300FixM12TanBeta10HiggsMassDist}. The simulations were carried out for $m_{0} = 4$ TeV and $\mathrm{tan}\beta =$ 10. From the results of \cite{bib:JEYounkinSPMartin}, we expected only a few models above our required lower limit of $m_{h} > 122$ GeV. Indeed, only $\sim$ 4\% of the simulated models achieve $m_{h} > 122$ GeV.\\
\indent Models that do not provide a Higgs boson mass with $122 < m_{h} < 128$ GeV must be rejected. So, the aim of this paper is to find a linear combination whose predictions with respect to the Higgs boson mass are more promising compared to that of Younkin et al.\\
\indent We investigated the linear combination \textbf{1} $\oplus$ \textbf{24} $\oplus$ \textbf{75} $\oplus$ \textbf{200}. To compare the results with those of \textbf{1} $\oplus$ \textbf{24} we also keep $m_{0} = 4$ TeV and $\mathrm{tan}\beta = 10$. We independently varied the mixing angles $\theta_{1}$, $\theta_{24}$, $\theta_{75}$ and $\theta_{200}$ in the range given in Section 2. In our model $\sim$ 36\% of the simulated models provide a Higgs boson with $m_{h} > 122$ GeV. Thus, a significant larger number of models in our parameterization can provide a Higgs boson in agreement with experimental measurements. This result is shown on the right hand side of Figure \ref{fig:su5Rep24And300FixM12TanBeta10HiggsMassDist}.

\begin{figure}[!ht]
  \begin{minipage}[t]{.48\textwidth}
    \begin{center}  
      \includegraphics[width=8.0cm]{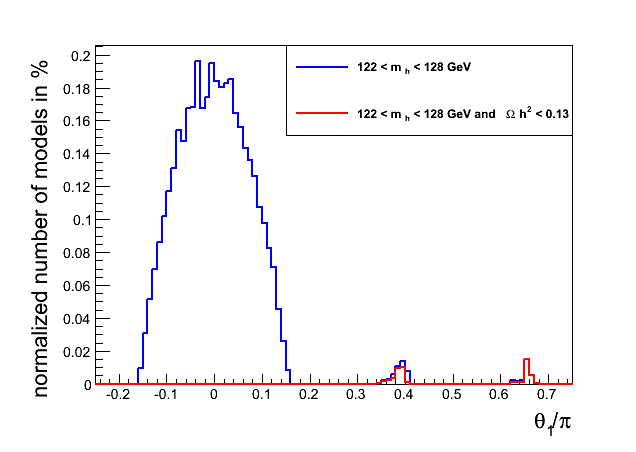}
    \end{center}
  \end{minipage}
  \hfill
  \begin{minipage}[t]{.48\textwidth}
    \begin{center}  
      \includegraphics[width=8.0cm]{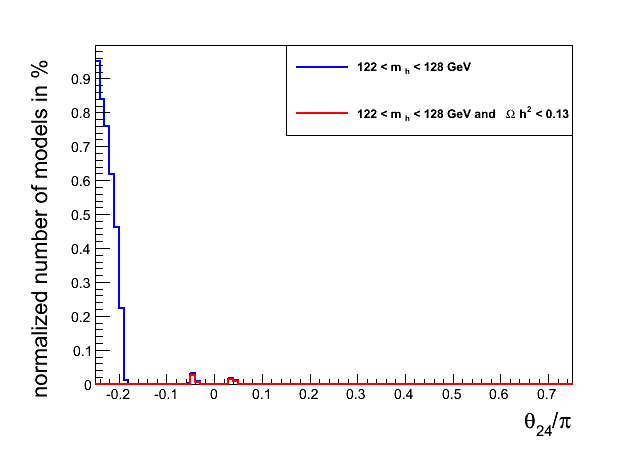}
    \end{center}
  \end{minipage}
  \caption[Higgs favoured regions in 2-dim. parameter space of $theta_{\mathrm{Representation}}$ for \textbf{1} $\oplus$ \textbf{24}]{Number of models fulfilling the Higgs mass constraint $122 < m_{h} < 128$ GeV (blue line) and the number of models fulfilling the Higgs mass constraint and the relic density constraint $\Omega h^{2} < 0.13$ (red line) is plotted versus the singlet mixing angle $\theta_{1}$ (left) and the 24-plet mixing angle $\theta_{24}$ (right). The number of models is normalized to the total number of simulated models}
  \label{fig:su5Rep24MHiggsAndOh2VsThetaI}
\end{figure}

\begin{figure}[!ht]
  \begin{minipage}[t]{.48\textwidth}
    \begin{center}  
      \includegraphics[width=8.0cm]{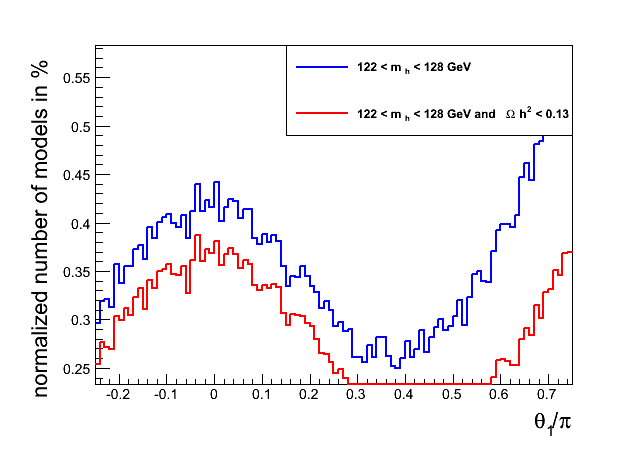}
    \end{center}
  \end{minipage}
  \hfill
  \begin{minipage}[t]{.48\textwidth}
    \begin{center}  
      \includegraphics[width=8.0cm]{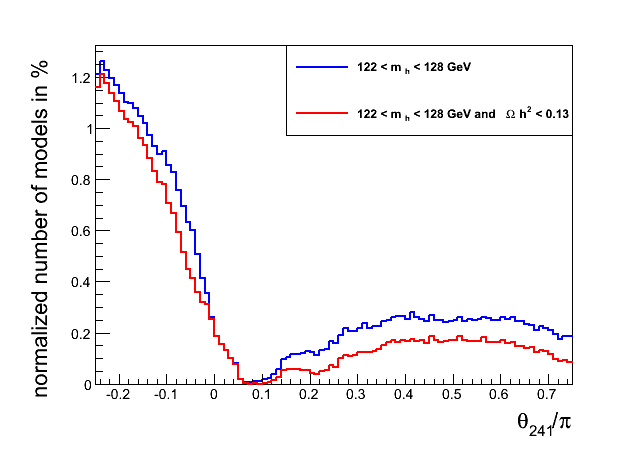}
    \end{center}
  \end{minipage}
  \begin{minipage}[t]{.48\textwidth}
    \begin{center}  
      \includegraphics[width=8.0cm]{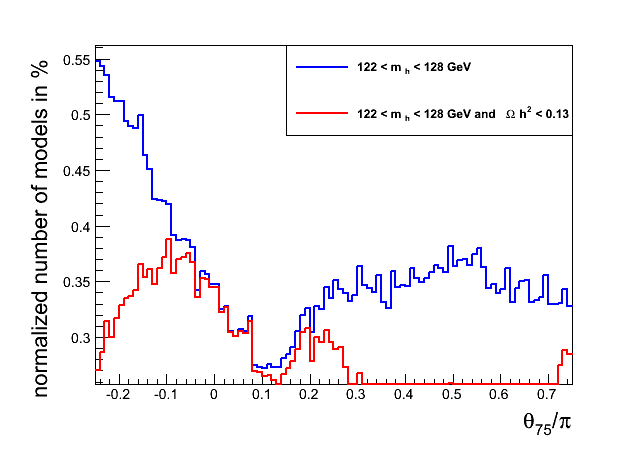}
    \end{center}
  \end{minipage}
  \hfill
  \begin{minipage}[t]{.48\textwidth}
    \begin{center}  
      \includegraphics[width=8.0cm]{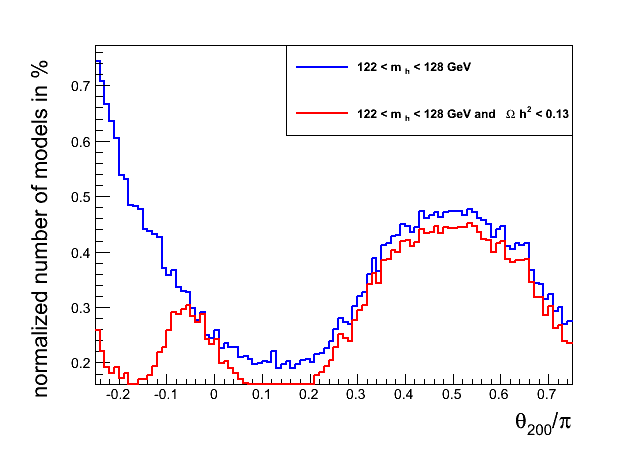}
    \end{center}
  \end{minipage}
  \caption[Higgs favoured regions in 4-dim. parameter space of $theta_{\mathrm{Representation}}$ for \textbf{1} $\oplus$ \textbf{24} $\oplus$ \textbf{75} $\oplus$ \textbf{200}]{Number of models with $122 < m_{h} < 128$ GeV (blue line) and $122 < m_{h} < 128$ GeV and $\Omega h^{2} < $0.13 (red line) of the four dimensional parameter space of \textbf{1} $\oplus$ \textbf{24} $\oplus$ \textbf{75} $\oplus$ \textbf{200}. The number of models fulfilling the Higgs mass constraint and relic density requirements is plotted versus the singlet mixing angle $\theta_{1}$ (upper left), the 24-plet mixing angle $\theta_{24}$ (upper right), the 75-plet mixing angle $\theta_{75}$ (lower left) and the 200-plet mixing angle $\theta_{200}$ (lower right). The number of models is normalized to the total number of simulated models}
  \label{fig:su5Rep300MHiggsAndOh2VsThetaI}
\end{figure}

\noindent In order to identify regions in the two (four) dimensional parameter space of mixing angles $\theta_{1}$ and $\theta_{24}$ for \textbf{1} $\oplus$ \textbf{24} ($\theta_{1}$, $\theta_{24}$, $\theta_{75}$ and $\theta_{200}$ for \textbf{1} $\oplus$ \textbf{24} $\oplus$ \textbf{75} $\oplus$ \textbf{200}), where $122 < m_{h} < 128$ GeV is given, we fixed the remaining input parameters of the model to the values given in the caption of Figure \ref{fig:su5Rep24And300FixM12TanBeta10HiggsMassDist} leaving only the two (four) angles as free parameters with $-0.25 < \theta_{i}/\pi < 0.75$. For \textbf{1} $\oplus$ \textbf{24} the number of models resulting in predictions $122 < m_{h} < 128$ GeV is plotted versus the mixing angle $\theta_{1}/\pi$ respectively $\theta_{24}/\pi$ in Figure \ref{fig:su5Rep24MHiggsAndOh2VsThetaI}. In Figure \ref{fig:su5Rep300MHiggsAndOh2VsThetaI} the number of models resulting in predictions $122 < m_{h} < 128$ GeV is plotted versus the mixing angles $\theta_{1}/\pi$, $\theta_{24}/\pi$, $\theta_{75}/\pi$ and $\theta_{200}/\pi$  (for \textbf{1} $\oplus$ \textbf{24} $\oplus$ \textbf{75} $\oplus$ \textbf{200}). Although the relic density is part of the discussion in the next Section, Figures \ref{fig:su5Rep24MHiggsAndOh2VsThetaI} and \ref{fig:su5Rep300MHiggsAndOh2VsThetaI} also show models that simultaneously fulfil $122 < m_{h} < 128$ GeV and $\Omega h^{2} < 0.13$ (red line).

\noindent In the case of \textbf{1} $\oplus$ \textbf{24} only restricted regions for $\theta_{1}/\pi$ and $\theta_{24}/\pi$ provide a Higgs boson with $122 < m_{h} < 128$ GeV. These are the regions with $-0.16 < \theta_{1}/\pi < 0.16$, $0.34 < \theta_{1}/\pi < 0.42$ and $0.62 < \theta_{1}/\pi < 0.68$ for the singlet contribution. The 24-plet contribution leads to $122 < m_{h} < 128$ for $\theta_{24} \lesssim -0.18$ and two small bumps at $\theta_{24}/\pi \sim -0.05$ and 0.03.\\

\indent The situation changes drastically when considering the most general linear combination \textbf{1} $\oplus$ \textbf{24} $\oplus$ \textbf{75} $\oplus$ \textbf{200}. Possible models that provide a Higgs boson with the correct mass can be obtained for all angles $\theta_{i}$ over the complete range $-0.25 < \theta_{i}/\pi < 0.75$ (Figure \ref{fig:su5Rep300MHiggsAndOh2VsThetaI}).\\
\indent As mentioned above, the red line in Figure \ref{fig:su5Rep24MHiggsAndOh2VsThetaI} and \ref{fig:su5Rep300MHiggsAndOh2VsThetaI} represents the number of models with $122 < m_{h} < 128$ GeV and $\Omega h^{2} < 0.13$. In the case of \textbf{1} $\oplus$ \textbf{24} regions with $122 < m_{h} < 128$ GeV and $\Omega h^{2}$ only partially coincide, such that in total only a few models provide a Higgs boson with the right mass and correct dark matter relic density.\\
\indent In the case of \textbf{1} $\oplus$ \textbf{24} $\oplus$ \textbf{75} $\oplus$ \textbf{200} over the complete range of the $\theta_{i}$'s models that simultaneously fulfil $\Omega h^{2} < 0.13$ and $122 < m_{h} < 128$ can be found.\\
\indent The relic density is part of the discussion in the next Section where we combine particle physics with astroparticle physics. There, we investigate our models predictions with respect to the relic density of the neutralino, which is assumed to be the dark matter particle.\\
\indent From now on we concentrate on linear combination \textbf{1} $\oplus$ \textbf{24} $\oplus$ \textbf{75} $\oplus$ \textbf{200}. For simplicity and graphical representation purposes we unify the mixing angles $\theta_{i}$, where $\theta_{1} = \theta_{24} = \theta_{75} = \theta_{200} \equiv \theta$.\\
\indent Of course, reducing the number of free parameters also reduces the allowed regions of the parameter space where the Higgs mass is consistent with measurements. But with this restriction still a factor of $\sim$ 10 more models in our parameterization survive the Higgs constraints compared to \textbf{1} $\oplus$ \textbf{24}.\\

\section{Relic Density for Non-universal Gaugino Masses}
\begin{figure}[!ht]
  \hfill
  \begin{minipage}[t]{.48\textwidth}
    \begin{center}  
      \includegraphics[width=8cm]{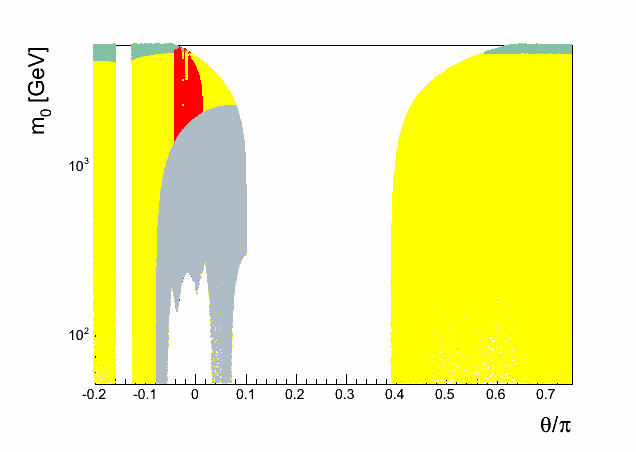}
    \end{center}
  \end{minipage}
  \hfill
  \begin{minipage}[t]{.48\textwidth}
    \begin{center}  
      \includegraphics[width=8cm]{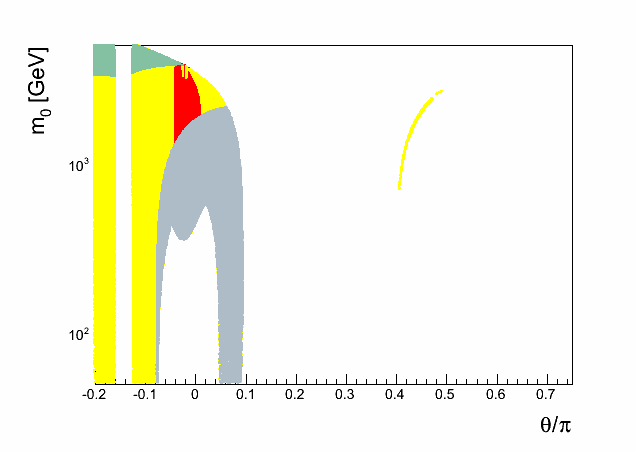}
    \end{center}
  \end{minipage}
  \hfill
  \caption[Calculated relic density for SU(5) representations \textbf{1} $\oplus$ \textbf{24} $\oplus$ \textbf{75} $\oplus$ \textbf{200}: $\Omega h^{2}$, $\mathrm{tan}\beta=10$ and $\mathrm{tan}\beta=45$]{Calculated relic density $\Omega h^{2}$ (color code) for SU(5) representations \textbf{1} $\oplus$ \textbf{24} $\oplus$ \textbf{75} $\oplus$ \textbf{200}, $\mathrm{tan}\beta=10$ (left) and $\mathrm{tan}\beta=45$ (right); green: prefered region where $\Omega h^{2} < 0.13$ and $122 < m_{h} < 128$ GeV; red: disfavored by cosmology with $\Omega h^{2} > 0.13$; yellow: $\Omega h^{2} < 0.13$, gray: excluded by $m_{squark} < 1.4$ TeV.}
  \label{fig:su5Rep300FixM12TanBeta10and45Oh2}
  \hfill
\end{figure}

The calculated relic density, $\Omega h^{2}$, predicted by the models introduced in Section 2, determines whether the corresponding model can give an explanation for dark matter. The relic density of dark matter, deduced from WMAP data, is restricted to $0.106 < \Omega h^{2} < 0.118$ given in \cite{bib:PdG}. We relaxed the previous constraint to $\Omega h^{2} < 0.13$. The relaxed upper constraint accounts for the possibility that R-Parity may not be conserved, while the relaxed lower bound includes the possibility that dark matter may not be only made up by one single particle. Furthermore, a lower value of $\Omega h^{2}$ with respect to WMAP data can be accepted if the missing relic density is filled up with dark matter particles produced non-thermally, e.g. the decay of long lived particles or cosmic strings \cite {bib:TMoroiLRandall}, \cite{bib:DChungEKolbARiotto}, \cite{bib:RJeannerotEtAl}.\\
\indent Throughout the following sections we present our results for $\mathrm{tan}\beta=$ 10 on the left and $\mathrm{tan}\beta=$ 45 on the right hand side. In Figure \ref{fig:su5Rep300FixM12TanBeta10and45Oh2} $\Omega h^{2}$ is shown for the mixing of representations \textbf{1} $\oplus$ \textbf{24} $\oplus$ \textbf{75} $\oplus$ \textbf{200}. All models with $\Omega h^{2} > 0.13$ are colored red, while models with $\Omega h^{2} < 0.13$ are colored yellow. Colored faint blue are parameter regions which yield squark masses $m_{squark} < 1.4$ TeV. Such squark masses are already excluded by LHC \cite{bib:ATLASSquarks}. Colored green regions refer to models with $\Omega h^{2} < 0.13$ and a Higgs mass of $122 < m_{h} < 128$ GeV. Green colored models are capable to account for dark matter and providing a Higgs boson with mass $122 < m_{h} < 128$.

\noindent The parameters in the white region between $0.1 \lesssim \theta/\pi \lesssim 0.4$ for both $\mathrm{tan}\beta$ values do not have a correct electro weak symmetry breaking, i.e. do not have a convergent $\mu$ from solving the renormalization group equations (RGE). The region between $-0.16 \lesssim \theta/\pi \lesssim -0.12$ for both $\mathrm{tan}\beta$ violate the LEP2 bound on the chargino mass \cite{bib:PdG}. The white region between $-0.07 \lesssim \theta/\pi \lesssim 0.04$ is forbidden due to tachyonic third generation sfermions ($\mathrm{tan}\beta =$ 45) or has a Higgs potential that is unbound from below or lead to charge and color breaking minima (see e.g. \cite{bib:FrereErAl} - \cite{bib:KounnasEtAl}), for $\mathrm{tan}\beta =$ 10. $\theta/\pi$ values greater than $\sim$ 0.4 for $\mathrm{tan}\beta =$ 45 lead to a tachyonic pseudoscalar Higgs boson $A$ and are excluded. For $\mathrm{tan}\beta =$ 10, several coannihilation regions occur. For $-0.1 \lesssim \theta/\pi \lesssim 0.08$ and $m_{0} \lesssim 1.5$ TeV as well as $0.54 < \theta/\pi < 0.6$ and $m_{0} < 400$ GeV the tau slepton, $\tilde{\tau}$, and the tau sneutrino, $\tilde{\nu}_{\tau}$, are nearly degenerate in their masses with the neutralino mass. Further, the top squark, $\tilde{t}$, coannihilates with the neutralino for $0.02 \lesssim \theta/\pi \lesssim 0.08$ and $m_{0} \lesssim 2$ TeV. Large parts of this coannihilation regions are excluded because the squarks are lighter than 1.4 TeV for $-0.05 < \theta/\pi < 0.12$ and $m_{0} < 2$ TeV (indicated in faint blue in Figure \ref{fig:su5Rep300FixM12TanBeta10and45Oh2}). Resonant annihilation regions with the pseudoscalar Higgs Boson $A$ and the lightest Higgs boson $h$ similar to h- and A- funnel regions of the cMSSM can be found in the vicinity of the LEP2 bound, for $h$, and $0.02 \lesssim \theta/\pi \lesssim 0.08$ and $m_{0} < 1$ TeV for $A$. The latter one is excluded by squarks that are too light.\ \\
\indent For $\mathrm{tan}\beta =$ 45, the above regions coincide, except the $\tilde{\tau}$ and $\tilde{\nu}_{\tau}$ coannihilation region for $\mathrm{tan}\beta \sim 0.5$, because of a tachyonic pseudoscalar Higgs boson. As for $\mathrm{tan}\beta =$ 10, most of the coannihilation regions are excluded by squarks that are too light for $-0.04 \lesssim \theta/\pi \lesssim 0.1$ and $m_{0} < 2$ TeV.\ \\
\indent The red region on the left side of Figure \ref{fig:su5Rep300FixM12TanBeta10and45Oh2} has a neutralino entirely made up of the bino similar to the bulk region of the cMSSM. The yellow regions are characterized by a neutralino, that is either wino dominated ($\theta/\pi < -0.05$ and $\theta/\pi > 0.58$) or higgsino dominated ($\theta/\pi > 0.04$ and $\theta/\pi < 0.52$) or a mixture between wino and higgsino ($0.02 \lesssim \theta/\pi \lesssim 0.06$ and $0.53 \lesssim \theta/\pi \lesssim 0.56$). The relic density in this regions is smaller than the lower bound deduced from WMAP data, i.e. smaller than 0.106. Moreover, in wino and/or higgsino dominated regions the lightest neutralino is almost degenerate with the lightest chargino and pairs of neutralinos annihilate very efficiently via t-channel chargino exchange into pairs of W bosons and the neutralinos relic population is depleted. Nevertheless, boundary conditions from WMAP correspond to thermally produced dark matter. Wino or higgsino like dark matter could have been produced non-thermally, in a sense that $\Omega h^{2}$ is decomposed into the sum of thermally plus non-thermally produced dark matter, $\Omega = \Omega_{\mathrm{therm}} + \Omega_{\mathrm{nontherm}}$ (see e.g. \cite{bib:RJeannerotEtAl}). The total relic density of cold dark matter can than be kept in agreement with observations. That is why we relax WMAP constraints to $0 < \Omega h^{2} < 0.13$.\\
\indent The same arguments apply for $\mathrm{tan}\beta =$ 45 (right hand side of Figure \ref{fig:su5Rep300FixM12TanBeta10and45Oh2}). Wino dominated regions are found for $\theta/\pi < -0.05$. For $\theta/\pi \gtrsim 0.04$ the neutralino is dominated by the higgsino and for $0.02 \lesssim \theta/\pi \lesssim 0.04$ the neutralino is a mixture between wino and higgsino. Again the red region is characterized by a bino dominated neutralino.\\
\indent Models, where $\mu$ has a desired small value, so that it can solve the little hierarchy problem coincide partially with higgsino dominated regions. Small $\mu$ regions are found as thin contours on top and on the right edge for both $\mathrm{tan}\beta =$ 10 and $\mathrm{tan}\beta =$ 45 of the allowed models. For $\mathrm{tan}\beta =$ 10 they are also found on top and on the left edge of allowed models. For $\mathrm{tan}\beta =$ 45 only a thin arc remains on the right side.\\
\indent After introducing the possible dark matter scenarios from the model setup mentioned in Section 2 we now investigate constraints given by selected indirect and direct detection experiments.

\section{Indirect Detection}
\begin{figure}[!ht]
  \hfill
  \begin{minipage}[t]{.48\textwidth}
    \begin{center}  
      \includegraphics[width=8cm]{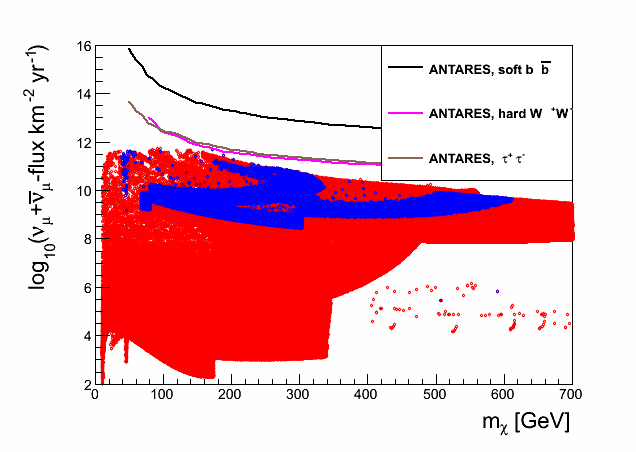}
    \end{center}
  \end{minipage}
  \hfill
  \begin{minipage}[t]{.48\textwidth}
    \begin{center}  
      \includegraphics[width=8cm]{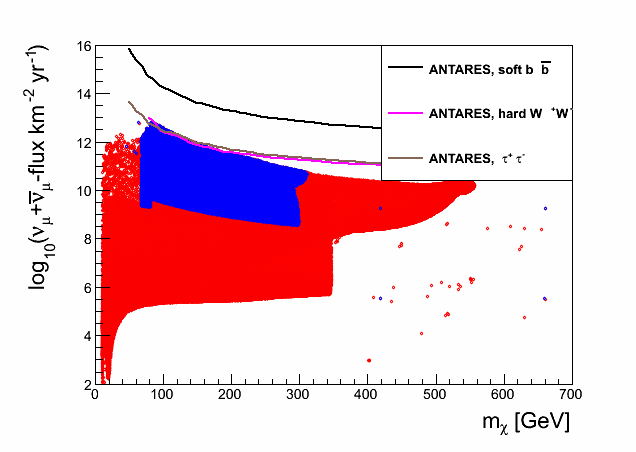}
    \end{center}
  \end{minipage}
  \caption[SU(5) representation \textbf{1} $\oplus$ \textbf{24} $\oplus$ \textbf{75} $\oplus$ \textbf{200}: $\nu_{\mu} - \bar{\nu}_{\mu}$ flux]{Sum of $\nu_{\mu}$ and $\bar{\nu}_{\mu}$ flux from dark matter annihilation for representation \textbf{1} $\oplus$ \textbf{24} $\oplus$ \textbf{75} $\oplus$ \textbf{200} of SU(5), $\mathrm{tan}\beta=10$ (left) and $\mathrm{tan}\beta=45$ (right); colors: blue: models with $\Omega h^{2} < 0.13$ and $122 < m_{h} < 128$ GeV; red: all other models; black line: ANTARES upper limit at 90\% C.L. \cite{bib:ZCharifVBertinPGay} on $\nu_{\mu} + \bar{\nu}_{\mu}$ flux for a 100\% annihilation into $b\bar{b}$; magenta line: annihilation into $W^{+}W^{-}$; brown line: annihilation into $\tau^{+}\tau^{-}$}
  \label{fig:su5Rep300FixM12TanBeta10and45NuFluxvsMChi}
  \hfill
\end{figure}

Although, there are plenty of indirect detection observables, e.g. photon flux, anti-proton flux, $e^{+}-e^{-}$ flux, that are worth being investigated, this paper is limited to the muon neutrino flux $\phi_{\mu_{\nu}}$ and the muon flux $\phi_{\mu}$ when talking about indirect detection.\ \\
\indent We focus on the indirect detection experiments ANTARES and IceCube. These experiments measure muons via the detection of Cerenkov light, which is emitted by the charged muons traveling through water or ice. The muon flux coming from below the detector is due to muon neutrinos which interact in charge current interactions close to the detector. In Figure \ref{fig:su5Rep300FixM12TanBeta10and45NuFluxvsMChi} the predicted integrated muon and anti muon neutrino flux from the Sun is plotted logarithmically versus the mass of the neutralino, $m_{\chi}$. The highest neutrino fluxes and therefore also highest muon fluxes are expected at the ''high higgsino'' or ''small $\mu$'' region explained in Section 3. In this parameter region the neutralinos can annihilate into Higgs and weak vector bosons resulting in a high muon neutrino flux. 

\begin{figure}[!ht]
  \hfill
  \begin{minipage}[t]{.48\textwidth}
    \begin{center}  
      \includegraphics[width=8cm]{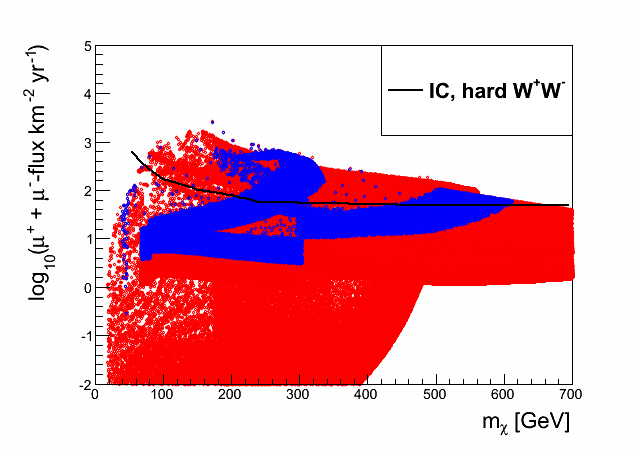}
    \end{center}
  \end{minipage}
  \hfill
  \begin{minipage}[t]{.48\textwidth}
    \begin{center}  
      \includegraphics[width=8cm]{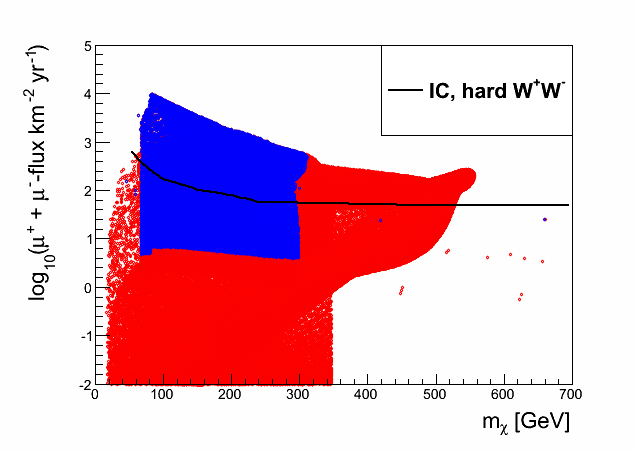}
    \end{center}
  \end{minipage}
  \caption[SU(5) representation \textbf{1} $\oplus$ \textbf{24} $\oplus$ \textbf{75} $\oplus$ \textbf{200}: $\mu^{+} - \mu^{-}$ flux]{Sum of $\mu^{+}$ and $\mu^{-}$ flux for representation \textbf{1} $\oplus$ \textbf{24} $\oplus$ \textbf{75} $\oplus$ \textbf{200}, $\mathrm{tan}\beta=10$ (left) and $\mathrm{tan}\beta=45$ (right), $m_{1/2} = 600$ GeV and $A_{0}=-600$ GeV; colors: blue: models with $\Omega h^{2} < 0.13$ and $122 < m_{h} < 128 GeV$; red: all other models; black line: IceCube upper limit at 90\% C.L. \cite{bib:IceCube} on $\mu$ flux for a 100\% annihilation into $W^{+}W^{-}$.}
  \label{fig:su5Rep300FixM12TanBeta10and45MuFluxvsMChi}
  \hfill
\end{figure}

\noindent The black, magenta and brown lines in Figure \ref{fig:su5Rep300FixM12TanBeta10and45NuFluxvsMChi} correspond to the ANTARES limit at 90\% Confidence Level assuming that all neutralinos annihilate exclusively into either $b\bar{b}$, $W^{+}W^{-}$ or $\tau^{+}\tau^{-}$, so that the limit is independent from the choice of the SUSY model. As can be seen in Figure \ref{fig:su5Rep300FixM12TanBeta10and45NuFluxvsMChi} ANTARES is not yet able to exclude the kind of models introduced in Section 2. The published ANTARES limit \cite{bib:ZCharifVBertinPGay} is based on 282.84 days of data taking that include a correction of 20\% for the 5 line detector configuration. More stringent limits are expected from the analysis of further data taken with the 12 line detector configuration.\\
\indent The predicted neutrino induced muon and anti muon fluxes from neutralino annihilations in the Sun are displayed in Figure \ref{fig:su5Rep300FixM12TanBeta10and45MuFluxvsMChi} for combination \textbf{1} $\oplus$ \textbf{24} $\oplus$ \textbf{75} $\oplus$ \textbf{200} and both $\mathrm{tan}\beta=$ 10 and $\mathrm{tan}\beta=$ 45. As in the case of the neutrino flux, the limits given in Figure \ref{fig:su5Rep300FixM12TanBeta10and45MuFluxvsMChi} assume a 100\% annihilation into $W^{+}W^{-}$. The gray line in Figure \ref{fig:su5Rep300FixM12TanBeta10and45MuFluxvsMChi} correspond to the limit at 90\% C.L. for the IceCube neutrino telescope \cite{bib:IceCube} with 86 strings including the low energy extension DeepCore (ICDC). From Figure \ref{fig:su5Rep300FixM12TanBeta10and45MuFluxvsMChi} it follows that still a reasonable amount of models, consistent with $\Omega h^{2} < 0.13$ and $122 < m_{h} < 128$, are not yet excluded by IceCube at 90\% C.L. for $\mathrm{tan}\beta =$ 10 as well as 45.\ \\

\begin{figure}[H]
  \hfill
  \begin{minipage}[t]{.48\textwidth}
    \begin{center}  
      \includegraphics[width=8cm]{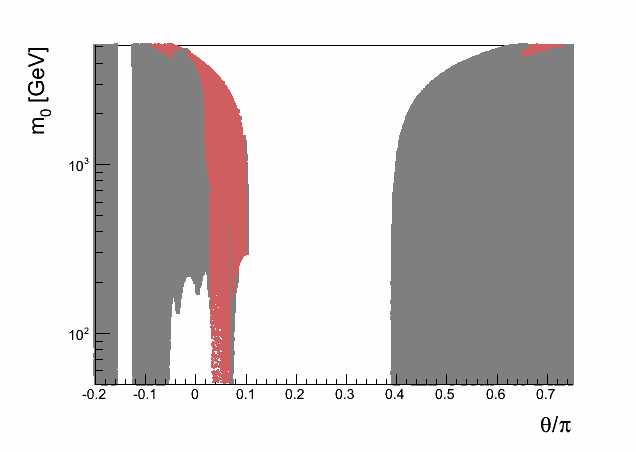}
    \end{center}
  \end{minipage}
  \hfill
  \begin{minipage}[t]{.48\textwidth}
    \begin{center}  
      \includegraphics[width=8cm]{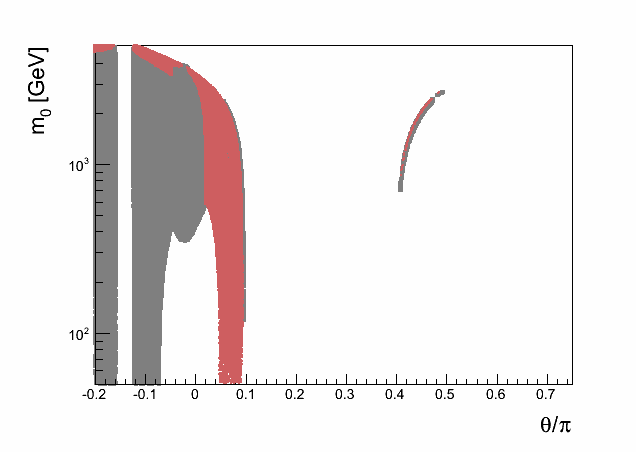}
    \end{center}
  \end{minipage}
  \caption[SU(5) representation \textbf{1} $\oplus$ \textbf{24} $\oplus$ \textbf{75} $\oplus$ \textbf{200}: ICDC exclusion from $W^{+}W^{-}$ channel for fix $M_{3}$, $\mathrm{tan}\beta=10$ and $\mathrm{tan}\beta=45$]{IceCube excluded models at 90\% C.L. for a 100\% annihilation into $W^{+}W^{-}$ for representation \textbf{1} $\oplus$ \textbf{24} $\oplus$ \textbf{75} $\oplus$ \textbf{200} of SU(5) $\mathrm{tan}\beta=10$ (left) and $\mathrm{tan}\beta=45$ (right); gray: not excluded, faint brown: excluded}
  \label{fig:su5Rep300FixM12TanBeta10and45IcdcExclHard}
  \hfill
\end{figure}

\noindent The exclusion limit of Figure \ref{fig:su5Rep300FixM12TanBeta10and45MuFluxvsMChi} is projected onto the $m_{0}-\theta$ plane to visualize which part of the parameter space can be excluded. These regions are shown in Figure \ref{fig:su5Rep300FixM12TanBeta10and45IcdcExclHard} for the annihilation channel $W^{+}W^{-}$.

\noindent For $\mathrm{tan}\beta =$ 10, large parts of the ''high higgsino'' or ''small $\mu$'' regions for $0 \lesssim \theta/\pi \lesssim 0.1$ and on top ($m_{0} \gtrsim 4$ TeV, $-0.2 \lesssim \theta/\pi \lesssim 0$ and $0.64 \lesssim \theta/\pi < 0.75$) can be excluded assuming all annihilations go into $W^{+}W^{-}$. This also applies for $\mathrm{tan}\beta =$ 45, where models with $m_{0} > 3.5$ TeV and $-0.2 < \theta/\pi \lesssim 0$ are excluded as well as models with $0 \lesssim \theta/\pi < 0.1$. Also, the thin arc on the right hand side for $\mathrm{tan}\beta =$ 45 is excluded by IceCube.\\
\indent From the indirect detections point of view the limits on the muon and anti muon flux given by the IceCube collaboration have the best power to exclude the kind of models we investigated.\\
\indent After investigation of exclusion capabilities of indirect detection with neutrino telescopes, in the next Section we focus on direct detection methods, i.e. elastic scattering interactions of a WIMP with a nucleon of the target material. We concentrate on results of the XENON 100 experiment \cite{bib:XENON100} and comment on the future extension XENON 1t \cite{bib:XENON1T}.

\section{Direct Detection}
\begin{figure}[!ht]
  \hfill
  \begin{minipage}[t]{.48\textwidth}
    \begin{center}  
      \includegraphics[width=8cm]{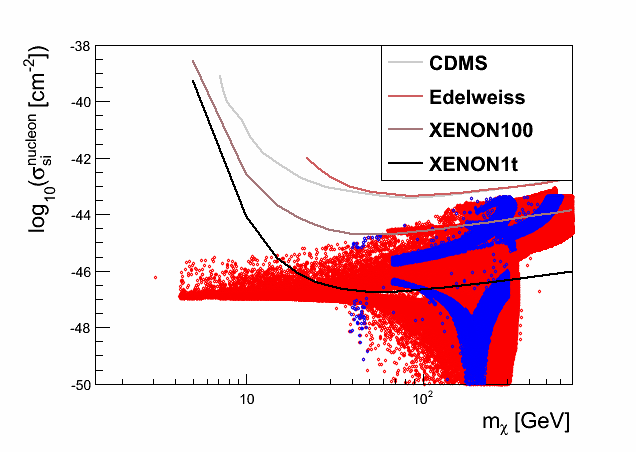}
    \end{center}
  \end{minipage}
  \hfill
  \begin{minipage}[t]{.48\textwidth}
    \begin{center}  
      \includegraphics[width=8cm]{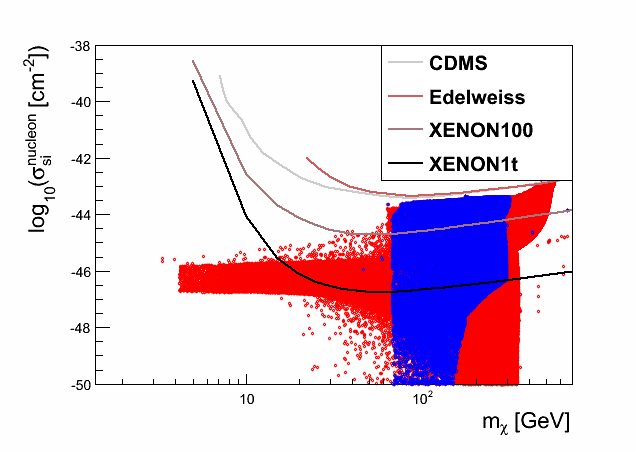}
    \end{center}
  \end{minipage}
  \caption[SU(5) representation \textbf{1} $\oplus$ \textbf{24} $\oplus$ \textbf{75} $\oplus$ \textbf{200}: $\sigma_{SI}^{nucleon}$ for fix $m_{1/2}$, $\mathrm{tan}\beta=10$ and $\mathrm{tan}\beta=45$]{Spin independent WIMP nucleon cross-section $\sigma_{SI}^{\mathrm{nucleon}}$ for representation \textbf{1} $\oplus$ \textbf{24} $\oplus$ \textbf{75} $\oplus$ \textbf{200} of SU(5), $\mathrm{tan}\beta=10$ (left) and $\mathrm{tan}\beta=45$ (right); blue: models with $\Omega h^{2} < 0.13$ and $122 < m_{h} < 128$ GeV; red: all other models; shown are 90\% C.L. limits from CDMS \cite{bib:CDMSII} (gray line), Edelweiss \cite{bib:Edelweiss} (brown line) and XENON 100 \cite{bib:XENON100} collaborations (shaded brown line) as well as the predicted limit for XENON 1t \cite{bib:XENON1T} detector (black line).}
  \label{fig:su5Rep300FixM12TanBeta10and45SigsiNucvsMChi}
  \hfill
\end{figure}

In this Section we focus on direct detection methods of dark matter, i.e. we compare the phenomenology of our models with the latest results given by the XENON collaboration. We concentrate on the spin independent WIMP nucleon cross-section $\sigma_{SI}^{\mathrm{nucleon}}$. The predicted cross-sections are  plotted in Figure \ref{fig:su5Rep300FixM12TanBeta10and45SigsiNucvsMChi} versus the mass of the neutralino $m_{\chi}$ which is assumed to be the WIMP.

\squeezeup

\begin{figure}[!ht]
  \hfill
  \begin{minipage}[t]{.48\textwidth}
    \begin{center}  
      \includegraphics[width=8cm]{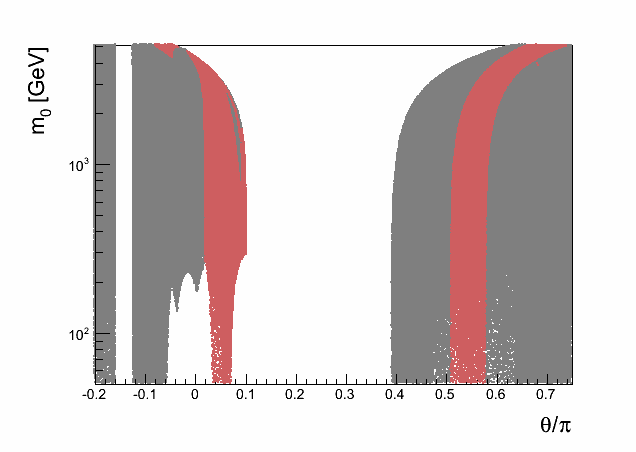}
    \end{center}
  \end{minipage}
  \hfill
  \begin{minipage}[t]{.48\textwidth}
    \begin{center}  
      \includegraphics[width=8cm]{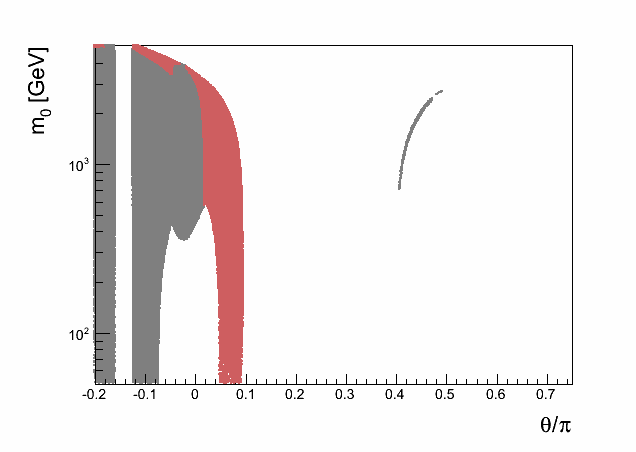}
    \end{center}
  \end{minipage}
  \caption[SU(5) representation \textbf{1} $\oplus$ \textbf{24} $\oplus$ \textbf{75} $\oplus$ \textbf{200}: XENON 100 exclusion regions for fix $m_{1/2}$, $\mathrm{tan}\beta=10$ and $\mathrm{tan}\beta=45$]{Excluded regions of the parameter space from XENON 100 for representation \textbf{1} $\oplus$ \textbf{24} $\oplus$ \textbf{75} $\oplus$ \textbf{200} of SU(5), $\mathrm{tan}\beta=10$ (left) and $\mathrm{tan}\beta=45$ (right); gray: not excluded; faint brown: excluded at 90\% C.L.}
  \label{fig:su5Rep300FixM12TanBeta10and45SigsiNucExclXENON100}
  \hfill
\end{figure}

\begin{figure}[!ht]
  \hfill
  \begin{minipage}[t]{.48\textwidth}
    \begin{center}  
      \includegraphics[width=8cm]{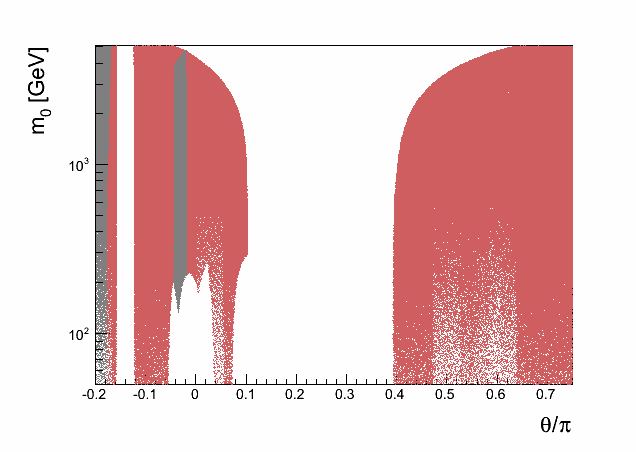}
    \end{center}
  \end{minipage}
  \hfill
  \begin{minipage}[t]{.48\textwidth}
    \begin{center}  
      \includegraphics[width=8cm]{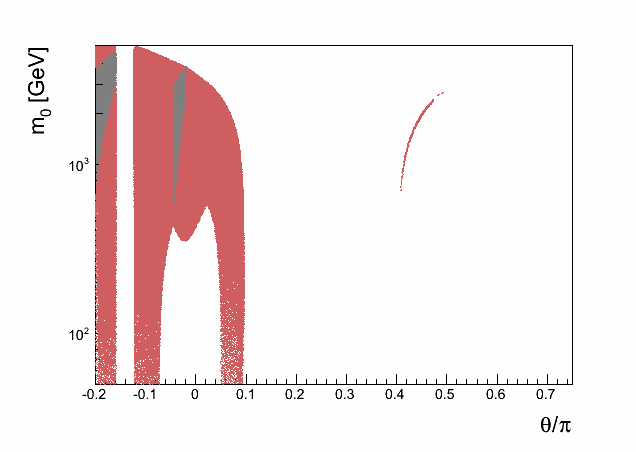}
    \end{center}
  \end{minipage}
  \caption[SU(5) representation \textbf{1} $\oplus$ \textbf{24} $\oplus$ \textbf{75} $\oplus$ \textbf{200}: XENON 1t exclusion regions for fix $m_{1/2}$, $\mathrm{tan}\beta=10$ and $\mathrm{tan}\beta=45$]{Excludable regions of the parameter space from XENON 1t for representation \textbf{1} $\oplus$ \textbf{24} $\oplus$ \textbf{75} $\oplus$ \textbf{200} of SU(5), $\mathrm{tan}\beta=10$ (left) and $\mathrm{tan}\beta=45$ (right); gray: not excludable; faint brown: excludable at 90\% C.L.}
  \label{fig:su5Rep300FixM12TanBeta10and45SigsiNucExclXENON1t}
  \hfill
\end{figure}

\noindent Clearly visible is the fact, that only XENON 100 of all direct detection experiments shown here is able to exclude any of the simulated models. Nevertheless, several of the simulated models with a Higgs mass of $122 < m_{h} < 128$ and $\Omega h^{2} < 0.13$, are not yet excluded by direct detection experiments.\\ 
\indent Even with the predicted sensitivity of the future extension XENON 1t (black line in Figure \ref{fig:su5Rep300FixM12TanBeta10and45SigsiNucvsMChi}) a reasonable number of models that fulfill our requirements for $\Omega h^{2}$ and the Higgs mass $m_{h}$ survive. This is not the case for the previously studied models \cite{bib:JEYounkinSPMartin}, where $\sim$ 95\% of these models for low $\mathrm{tan}\beta$ and all models for high $\mathrm{tan}\beta$ are excludable by XENON 1t.\\
\indent The corresponding excluded (excludable) parameter space for XENON 100 (XENON 1t) in the $m_{0}-\theta$ plane is shown in Figure \ref{fig:su5Rep300FixM12TanBeta10and45SigsiNucExclXENON100} and \ref{fig:su5Rep300FixM12TanBeta10and45SigsiNucExclXENON1t}.

\indent For $\mathrm{tan}\beta =$ 10 (left hand side of Figure \ref{fig:su5Rep300FixM12TanBeta10and45SigsiNucExclXENON100}), XENON 100 excludes many of the ''small $\mu$/high higgsino'' models on the right and upper edge of the left ''island''. The faint red band on the right ''island'' that is excluded by XENON 100 corresponds to models where the neutralino is a mixture between wino and higgsino. In this region $\mu$ is already too high to solve fine-tuning problems, as it is the case for small $\mu$.\\
\indent For $\mathrm{tan}\beta =$ 45 (right hand side of Figure \ref{fig:su5Rep300FixM12TanBeta10and45SigsiNucExclXENON100}), again the right and upper edge of the left ''island'' is excluded by XENON 100. As for the case of low $\mathrm{tan}\beta$ models belonging to that region have a small $\mu$ and solve the little hierarchy problem of supersymmetry. Complementary to IceCube, no models on the thin arc on the right hand side are excluded by XENON 100.\\
\indent Almost the whole cMSSM like focus point region with a small $\mu$ and a large higgsino fraction in the neutralinos composition can be tested by XENON 1t for $\mathrm{tan}\beta=$ 10 and 45 (Figure \ref{fig:su5Rep300FixM12TanBeta10and45SigsiNucExclXENON1t}). Only a few models for $-0.04 \lesssim \theta/\pi < 0$ and $m_{0} > 4$ TeV ($\mathrm{tan}\beta =$ 10) and $m_{0} > 3.5$ TeV ($\mathrm{tan}\beta =$ 45) may survive. In the case $\mathrm{tan}\beta=$ 10, all models on the right ''island'' that fulfill $\Omega h^{2} < 0.13$ and $122 < m_{h} < 128$ GeV would be excludable by XENON 1t. Only models with $\theta/\pi \lesssim 0.18$ still provide a consistent Higgs mass with the relic density requirement. For $\mathrm{tan}\beta =$ 45 a triangular shaped region survives the predicted sensitivity of XENON 1t. It can be found for $m_{0} > 3.4$ TeV and $\theta < 0.16$.\\
\indent When comparing the experimental limits presented in Section 5 and 6 to the model predictions one should keep in mind the uncertainties and approximations which are contained in the models. These uncertainties comprise the WMIP nucleon cross-section, $\sigma_{SI}^{\mathrm{nucleon}}$, from uncertainties in nuclear matrix elements and uncertainties in the local WIMP density, which affect the capture rate of neutralinos in the Sun and also the predictability with respect to direct detection.

\section{Combined Excluded Regions of the Parameter Space}
In this section we summarize the number of models simulated, models that fulfill our relic density requirement and models that can be excluded by ANTARES for $W^{+}W^{-}$, $\tau^{+}\tau^{-}$, IceCube + DeepCore (ICDC) for $W^{+}W^{-}$ and XENON with and without respect to a correct relic density. Numbers can be found in Table \ref{tbl:exclStatisticsSummary}. The prefix ''DM'' in Table \ref{tbl:exclStatisticsSummary} means models that have relic density with $\Omega h^{2} < 0.13$ while the prefix ''Higgs'' are those with $122 < m_{h} < 128$ GeV.\\  
\indent We present excluded regions of the parameter space when combining the most stringent constraints coming from the Higgs boson as well as XENON 100 and IceCube limits.

\begin{table}[!ht]
 \begin{center}
 \begin{tabular}{l|r|r|}
  \cline{2-3}
  & \multicolumn{2}{|c|}{\textbf{1} $\oplus$ \textbf{24} $\oplus$ \textbf{75} $\oplus$ \textbf{200}} \\
  \cline{2-3}
  & $\mathrm{tan}\beta$=10 & $\mathrm{tan}\beta=$45 \\
  \hline
  \multicolumn{1}{|l|}{$n_{\mathrm{models}}$ simulated} & 4 335 000 & 4 335 000 \\
  \hline
  \multicolumn{1}{|l|}{$n_{\mathrm{models}}$ physical} & 1 911 336 & 1 026 480 \\
  \hline
  \multicolumn{1}{|l|}{$n_{\mathrm{models}}$ dark matter (DM)} & 1 586 452 & 728 981 \\
  \hline
  \multicolumn{1}{|l|}{$n_{\mathrm{models}}$ ANTARES excl. $W^{+}W^{-}$} & 0 & 793 \\
  \hline
  \multicolumn{1}{|l|}{$n_{\mathrm{models}}$ ANTARES excl. $\tau^{+}\tau^{-}$} & 0 & 303 \\
  \hline
  \multicolumn{1}{|l|}{$n_{\mathrm{models}}$ ICDC excl. $W^{+}W^{-}$} & 155 086 & 189 527 \\
  \hline
  \multicolumn{1}{|l|}{$n_{\mathrm{models}}$ XENON 100 excluded} & 206 132 & 188 897 \\
  \hline
  \multicolumn{1}{|l|}{$n_{\mathrm{models}}$ XENON 1t excludable} & 1 583 595 & 781 573 \\
  \hline
  \multicolumn{1}{|l|}{$n_{\mathrm{models}}$ DM ANTARES excl. $W^{+}W^{-}$} & 0 & 0 \\
  \hline
  \multicolumn{1}{|l|}{$n_{\mathrm{models}}$ DM ANTARES excl. $\tau^{+}\tau^{-}$} & 0 & 303 \\
  \hline
  \multicolumn{1}{|l|}{$n_{\mathrm{models}}$ DM ICDC excl. $W^{+}W^{-}$} & 153 595 & 187 947 \\
  \hline
  \multicolumn{1}{|l|}{$n_{\mathrm{models}}$ DM XENON 100 excluded} & 204 978 & 187 610 \\
  \hline
  \multicolumn{1}{|l|}{$n_{\mathrm{models}}$ DM XENON 1t excludable} & 1 473 722 & 648 352 \\
  \hline
  \multicolumn{1}{|l|}{$n_{\mathrm{models}}$ Higgs} & 51 217 & 56 835 \\
  \hline
  \multicolumn{1}{|l|}{$n_{\mathrm{models}}$ Higgs DM} & 49 170 & 56 624 \\
  \hline
  \multicolumn{1}{|l|}{$n_{\mathrm{models}}$ Higgs ANTARES excl. $W^{+}W^{-}$} & 0 & 793 \\
  \hline
  \multicolumn{1}{|l|}{$n_{\mathrm{models}}$ Higgs ANTARES excl. $\tau^{+}\tau^{-}$} & 0 & 303 \\
  \hline
  \multicolumn{1}{|l|}{$n_{\mathrm{models}}$ Higgs ICDC excl. $W^{+}W^{-}$} & 9 712 & 24 886 \\
  \hline
  \multicolumn{1}{|l|}{$n_{\mathrm{models}}$ Higgs XENON 100 excluded} & 10 571 & 16 547 \\
  \hline
  \multicolumn{1}{|l|}{$n_{\mathrm{models}}$ Higgs XENON 1t excludable} & 38 512 & 45 949 \\
  \hline
 \end{tabular}
 \end{center}
 \caption{Summary Table of excluded/excludable models by the indirect detection experiments ANTARES and IceCube (ICDC) as well as the direct detection experiments XENON 100 and XENON 1t. The prefix ''DM'' means models with a relic density of $\Omega h^{2} < 0.13$, Higgs consistent means $122 < m_{h} < 128$ GeV.}
 \label{tbl:exclStatisticsSummary}
\end{table}

\begin{figure}[!ht]
  \hfill
  \begin{minipage}[t]{.48\textwidth}
    \begin{center}  
      \includegraphics[width=8cm]{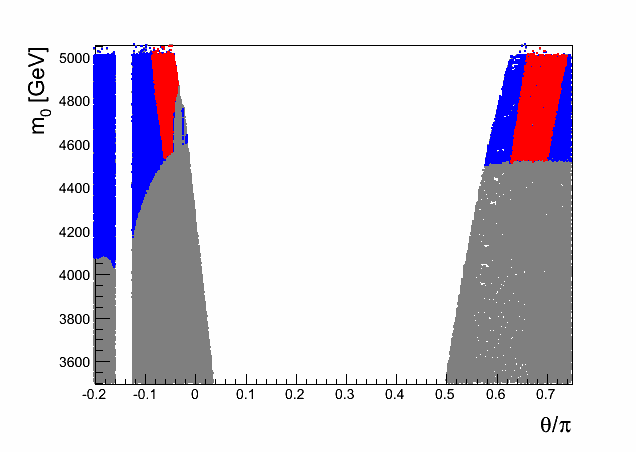}
    \end{center}
  \end{minipage}
  \hfill
  \begin{minipage}[t]{.48\textwidth}
    \begin{center}  
      \includegraphics[width=8cm]{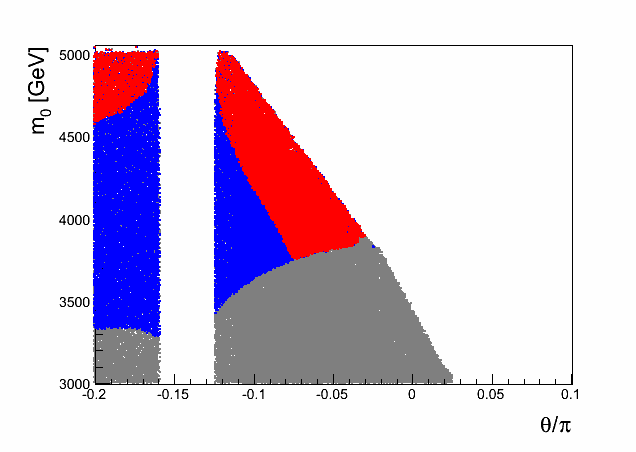}
    \end{center}
  \end{minipage}
  \caption[SU(5) representation \textbf{1} $\oplus$ \textbf{24} $\oplus$ \textbf{75} $\oplus$ \textbf{200}: combined IceCube and XENON 100 excluded regions for fix $m_{1/2}$, $\mathrm{tan}\beta=10$ and $\mathrm{tan}\beta=45$]{Combined IceCube (annihilation channel $W^{+}W^{-}$) and XENON 100 excluded regions of the parameter space for representation \textbf{1} $\oplus$ \textbf{24} $\oplus$ \textbf{75} $\oplus$ \textbf{200} of SU(5), $\mathrm{tan}\beta=10$ (left) and $\mathrm{tan}\beta=45$ (right); blue: $122 < m_{h} < 128$ GeV and $\Omega h^{2} < 0.13$; red: IceCube excluded at 90\% C.L. or XENON 100 excluded at 90\% C.L. and $122 < m_{h} < 128$ GeV plus $0 < \Omega h^{2} < 0.13$; gray: $m_{h} < 122$ GeV or $m_{h} > 128$ GeV or $\Omega h^{2} > 0.13$.}
  \label{fig:su5Rep300FixM12TanBeta10and45Xenon100ICExclWMAPHiggs}
  \hfill
\end{figure}

\noindent The allowed regions (colored blue) that fulfill the constraints $\Omega h^{2} < 0.13$ and $122 < m_{h} < 128$ GeV are summarized in Figure \ref{fig:su5Rep300FixM12TanBeta10and45Xenon100ICExclWMAPHiggs} (for the parameter space of representation \textbf{1} $\oplus$ \textbf{24} $\oplus$ \textbf{75} $\oplus$ \textbf{200}). Colored red are those models, that have a relic density $\Omega h^{2} < 0.13$ and Higgs boson with $122 < m_{h} < 128$ GeV and are excluded by either IceCube or XENON 100 at 90\% C.L. Models not satisfying $122 < m_{h} < 128$ GeV or $\Omega h^{2} < 0.13$ are colored gray.

\section{Variable Gaugino Mass Ratios}
In the previous Sections we have shown, that a correct Higgs mass can be achieved for models with fixed mixing angles $\theta_{i} \equiv \theta$ and $m_{0} \gtrsim 4$ TeV. Thus, the linear combination \textbf{1} $\oplus$ \textbf{24} $\oplus$ \textbf{75} $\oplus$ \textbf{200} provides significantly more models that fulfil the constraint $122 < m_{h} < 128$ GeV compared to \textbf{1} $\oplus$ \textbf{24}.\\
\indent Nevertheless, we had to relax the relic density requirements and include the possibility for dark matter to be produced non-thermally. Otherwise, most of the simulated models do not produce a sufficient amount of thermally produced dark matter.\\
\indent In this Section we present a preliminary study of deviations from the mass ratios of SU(5) representations. We parameterize the gaugino masses $M_{1}$, $M_{2}$ and $M_{3}$ similar to Equation \ref{eq:spmparaGeneral}, given by

\squeezeup

\begin{align}\label{eq:spmparaVarCoeffs}
 M_{1} & =  m_{1/2} \left( \mathrm{cos}\left( \theta \right) + a \ \mathrm{sin}\left( \theta \right) \right) \nonumber \\
 M_{2} & =  m_{1/2} \left( \mathrm{cos}\left( \theta \right) + b \ \mathrm{sin}\left( \theta \right) \right) \nonumber \\
 M_{3} & =  m_{1/2} \left( \mathrm{cos}\left( \theta \right) + c \ \mathrm{sin}\left( \theta \right) \right) 
\end{align}

\noindent where coefficients a, b and c are allowed to vary in the range [-10,10]. We fixed $m_{0} = $ 3 TeV, $m_{1/2} = $ 600 GeV and $A_{0} = -m_{1/2}$. We simulated models for $\theta$ and $\mathrm{tan}\beta$ pairs $(\theta,\mathrm{tan}\beta)$ = (-45,10), (-45,45), (115,10) and (115,45). The number of models simulated, as well as the number of models that are consistent with $122 < m_{h} < 128$ GeV and $0.09 < \Omega h^{2} < 0.13$ are listed in Table \ref{tbl:varCoeffNumberOfModels}. We required a more stringent bound on the relic density $0.09 < \Omega h^{2} < 0.13$, in contrast to the previous sections.

\begin{table}[h]
 \vspace{0.5cm}
  \begin{center}
  \begin{tabular}{|l|l|l|}
    \hline
    \backslashbox{$\bm{\theta}$}{$\bm{\mathrm{tan}\beta}$} & \multicolumn{1}{|c|}{\textbf{10}} & \multicolumn{1}{|c|}{\textbf{45}} \\
    \hline
    \multirow{3}{*}{$\bm{-45^{\circ}}$} & $n_{\mathrm{models}}$ simulated: 599 470 & $n_{\mathrm{models}}$ simulated: 411 194 \\
    & $n_{\mathrm{models}}$ Higgs: 429 612 & $n_{\mathrm{models}}$ Higgs: 367 354 \\
    & $n_{\mathrm{models}}$ DM: 10 537 & $n_{\mathrm{models}}$ DM: 6 037 \\
    & $n_{\mathrm{models}}$ Higgs + DM: 3 673 & $n_{\mathrm{models}}$ Higgs + DM: 5 147\\ 
    \hline
    \multirow{3}{*}{$\bm{115^{\circ}}$} & $n_{\mathrm{models}}$ simulated: 551 683 & $n_{\mathrm{models}}$ simulated: 340 497 \\
    & $n_{\mathrm{models}}$ Higgs: 443 992 & $n_{\mathrm{models}}$ Higgs: 314 617 \\
    & $n_{\mathrm{models}}$ DM: 14 444 & $n_{\mathrm{models}}$ DM: 8 243 \\
    & $n_{\mathrm{models}}$ Higgs + DM: 9 045 & $n_{\mathrm{models}}$ Higgs + DM: 7 654\\
    \hline
  \end{tabular}
  \end{center}
  \caption{Summary Table of the number of simulated models, the number of models with $122 < m_{h} < 128$ GeV (labeled $n_{\mathrm{models}}$ Higgs), the number of models with $0.09 < \Omega h^{2} < 0.13$ (labeled $n_{\mathrm{models}}$ DM) and the number of models with $122 < m_{h} < 128$ GeV and additionally $0.09 < \Omega h^{2} < 0.13$ (labeled $n_{\mathrm{models}}$ Higgs + DM).}
  \label{tbl:varCoeffNumberOfModels}
\end{table}

\noindent Models with $122 < m_{h} < 128$ GeV are labeled ''$n_{\mathrm{models}}$ Higgs'' in Table \ref{tbl:varCoeffNumberOfModels}. Models with relic densities $0.09 < \Omega h^{2} < 0.13$ are labeled ''$n_{\mathrm{models}}$ DM''. The constraint on $\Omega h^{2}$ slightly deviates from the limits given by the WMAP collaboration. It includes the possibility of a broken R-parity and thus decaying dark matter as well as an additional dark matter component, e.g. axions.\\
\indent To find the optimal triple (a,b,c) for each of the pairs ($\theta$,$\mathrm{tan}\beta$) from Table \ref{tbl:varCoeffNumberOfModels} that describes $m_{h}$ and $\Omega h^{2}$ best, we performed a $\chi^{2}$ analysis according to

\begin{equation}
  \chi^{2} = \chi^{2}_{\mathrm{Higgs}} + \chi^{2}_{\Omega h^{2}} = \frac{ (m_{h,\mathrm{predicted}} - m_{h, \mathrm{observed}})^{2} }{ (\sigma_{\mathrm{Higgs}}^{\mathrm{observed}})^{2} + (\sigma_{\mathrm{Higgs}}^{\mathrm{theo}})^{2} } + \frac{ (\Omega h^{2}_{\mathrm{predicted}} - \Omega h^{2}_{\mathrm{observed}})^{2} }{ (\sigma_{\Omega h^{2}}^{\mathrm{observed}})^{2} }
\end{equation}

\noindent where we took $\Omega h^{2}_{\mathrm{observed}} = 0.11 \pm 0.02$. $m_{h,\mathrm{observed}} = 125.3$ GeV is the observed mass of the Higgs boson given by the CMS collaboration \cite{bib:CMSHiggs}. $\sigma_{\mathrm{Higgs}}^{\mathrm{theo}} =  3$ GeV and $\sigma_{\mathrm{Higgs}}^{\mathrm{observed}} =  0.4 (\mathrm{stat.}) + 0.5 (\mathrm{syst.})$ GeV are the theoretical and experimental uncertainties, respectively on the Higgs boson. The resulting $\chi^{2}$-values for (a,b,c) are listed in Table \ref{tbl:varCoeffChi2Analysis}

\begin{table}[h]
 \vspace{0.5cm}
  \begin{center}
  \begin{tabular}{|l|l|l|}
    \hline
    \backslashbox{$\bm{\theta}$}{$\bm{\mathrm{tan}\beta}$} & \multicolumn{1}{|c|}{\textbf{10}} & \multicolumn{1}{|c|}{\textbf{45}} \\
    \hline
    \multirow{6}{*}{$\bm{-45^{\circ}}$} & $\chi^{2} = 5\cdot 10^{-4}$ & $\chi^{2} = 4\cdot 10^{-4}$ \\
    & $m_{h} =$ 125.37 GeV & $m_{h} =$ 125.32 GeV \\
    & $\Omega h^{2} =$ 0.110 & $\Omega h^{2} =$ 0.110 \\
    & $a =$ -9.79 & $a =$ -6.48 \\
    & $b =$ -4.67 & $b =$ -3.07 \\
    & $c =$ -8.15 & $c =$ -7.03 \\
    \hline
    \multirow{6}{*}{$\bm{115^{\circ}}$} & $\chi^{2} = 3.15\cdot 10^{-4}$ & $\chi^{2} = 7.82 \cdot 10^{-5}$ \\
    & $m_{h} =$ 125.30 GeV & $m_{h} =$ 125.28 GeV \\
    & $\Omega h^{2} =$ 0.110 & $\Omega h^{2} =$ 0.110 \\
    & $a =$ 9.47 & $a =$ 7.21 \\
    & $b =$ 5.01 & $b =$ -2.88 \\
    & $c =$ 7.45 & $c =$ 7.75 \\
    \hline
  \end{tabular}
  \end{center}
  \caption{Model parameters (a,b,c) resulting from the $\chi^{2}$-analysis for the combinations ($\theta$,$\mathrm{tan}\beta$) listed in Table \ref{tbl:varCoeffNumberOfModels}}
  \label{tbl:varCoeffChi2Analysis}
\end{table}

\noindent We use the triples (a,b,c) from the four simulated nodes listed in Table \ref{tbl:varCoeffChi2Analysis} to fit linear functions a(x,y), b(x,y) and c(x,y) for the coefficients given in Eq. \ref{eq:spmparaVarCoeffs}, where $x = \mathrm{sin}(\theta)$ and $y = \mathrm{tan}\beta$. These linear functions are given by

\squeezeup

\begin{eqnarray}\label{eq:linearizedCoeffs}
  a & = & a(x,y) = -1.60 + 12.9x + 0.02y - 0.10xy \nonumber \\
  b & = & b(x,y) = 0.31 + 7.68x - 0.07y - 0.17xy \nonumber \\
  c & = & c(x,y) = -1.53 + 9.82x + 0.02y - 0.01xy 
\end{eqnarray}

\noindent With these linearized coefficients we simulated $\sim$ 300 000 models with our benchmark point input parameters $m_{0} = 3$ TeV, $m_{1/2} = 600$ GeV and $A_{0} = -m_{1/2}$. We varied $\theta$ from -45 to 135 degree and $\mathrm{tan}\beta$ from 2 to 60.\\

\begin{figure}[!ht]
  \hfill
  \begin{minipage}[t]{.48\textwidth}
    \begin{center}  
      \includegraphics[width=8.0cm]{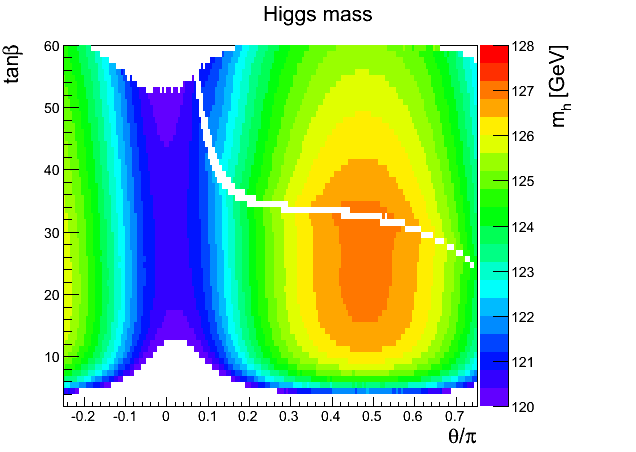}
    \end{center}
  \end{minipage}
  \hfill
  \begin{minipage}[t]{.48\textwidth}
    \begin{center}  
      \includegraphics[width=8.0cm]{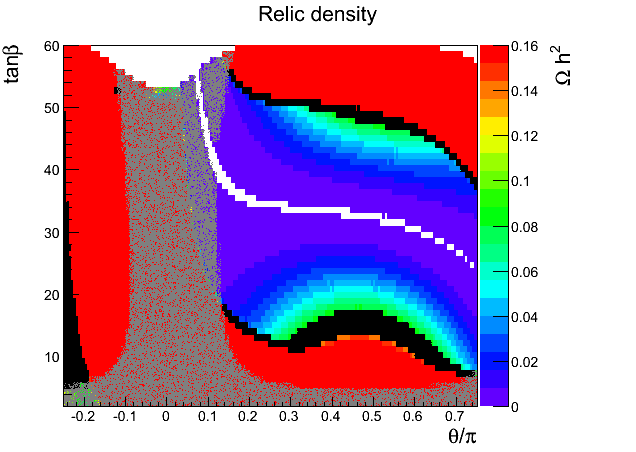}
    \end{center}
  \end{minipage}
  \caption[$m_{h}$ and $\Omega h^{2}$ for linearized coefficients]{Predicted Higgs mass (left) and relic density (right) for linearized coefficients a(x,y), b(x,y) and c(x,y) in the $\theta/\pi$-$\mathrm{tan}\beta$ plane; values for $m_{h}$ are color coded from 120 GeV (purple) to 128 GeV (red); values for $\Omega h^{2}$ are color coded from 0 (purple) to 0.16 (red). The curly white band in the middle ($0.08 \lesssim \theta/\pi < 0.75$ and $26 < \mathrm{tan}\beta < 55$) is excluded by the LEP2 limit on the chargino mass. Solid black regions represent models that simultaneously fulfil $122 < m_{h} < 128 GeV$ and $0.09 < \Omega h^{2} < 0.13$, grey shaded regions have $m_{h} <$ 122 GeV or $m_{h} >$ 128 GeV.}
  \label{fig:linearizedCoeffsMHiggsOh2}
  \hfill
\end{figure}

\squeezeup

\indent Only for $-0.1 \lesssim \theta/\pi \lesssim 0.1$ $m_{h}$ is below 122 GeV (Figure \ref{fig:linearizedCoeffsMHiggsOh2}). As for these small angles $\mathrm{sin}(x) \sim x$ and $\mathrm{cos}(x) \sim$ 1, the region $-0.1 \lesssim \theta/\pi \lesssim 0.1$ is obviously dominated by the singlet and unified gaugino masses like in cMSSM. This ''cMSSM'' like Higgs region is already known to be disfavored by the measurements of CMS and ATLAS. Further, small values of $\mathrm{tan}\beta$, i.e. $\mathrm{tan}\beta \lesssim 5$ do not satisfy the lower bound $m_{h} > 122$ GeV. Further, the curly white band in the middle of the plot  is excluded by the LEP2 limit on the chargino mass \cite{bib:PdG}.\\
\indent Large parts of the parameter space in the $\theta$-$\mathrm{tan}\beta$ plane have $\Omega h^{2} > 0.16$ (Figure \ref{fig:linearizedCoeffsMHiggsOh2}). Two extended regions where $\Omega h^{2} \lesssim 0.13$ and $122 < m_{h} < 128$ GeV is fulfilled can be found for $5 \lesssim \mathrm{tan}\beta \lesssim 32$ and $\theta/\pi < -0.16$, $8 \lesssim \mathrm{tan}\beta \lesssim 54$ and $0.1 \lesssim \theta/\pi 0.75$. These regions are characterized by a neutralino that is a pure wino. Further, the neutralino is nearly degenerate in its mass with the lightest chargino and pairs of neutralinos can annihilate into pairs of W bosons via t-channel chargino exchange. Similar regions were already found in Section 4. Unfortunately, this annihilation process is very efficient and the relic density is pushed below 0.09 (light green/turky and purple regions in Figure \ref{fig:linearizedCoeffsMHiggsOh2}). Like in Section 4, where we allowed dark matter to be produced non-thermally, this assumption has to be made to make models in this regions viable dark matter models. Solid black colored regions on the right hand side of Figure \ref{fig:linearizedCoeffsMHiggsOh2} fulfil $0.09 < \Omega h^{2} < 0.13$ and $122 < m_{h} < 128$ GeV. There, the wino neutralino is heavy enough ($\mathcal{O}$(2 TeV)) to produce the right amount of thermally produced dark matter \cite{bib:JDWells}.\\
\indent Summarizing the above results: we simulated approximately 300 000 models. $\sim$ 73\% of them are consistent with $122 < m_{h} < 128$ GeV. $\sim$ 4\% of all models have a relic density with $0.09 < \Omega h^{2} < 0.13$. 91\% of models with $0.09 < \Omega h^{2} < 0.13$ provide a Higgs boson with a mass consistent with measurements. Of course, relaxing the relic density bound to $\Omega h^{2} < 0.13$ (see Section 4) increases the number of models consistent with $\Omega h^{2}$. In that case $\sim$ 45\% of all models fulfil the constraint on $\Omega h^{2}$. $\sim$ 90\% of these models have a Higgs boson with mass $122 < m_{h} < 128$ GeV.\\
\indent We imposed a linear dependence of coefficients (a,b,c) on the mixing angle $\mathrm{\theta}$ and $\mathrm{tan}\beta$. Simulating more nodes for pairs $(\theta,\mathrm{tan}\beta)$ would allow to introduce coefficients (a,b,c) that depend non-linearly on $(\theta,\mathrm{tan}\beta)$, allowing broader regions in the $\theta$-$\mathrm{tan}\beta$ plane that fulfil $0.09 < \Omega h^{2} < 0.13$.\\
\indent Nevertheless, this preliminary analysis indicates, that variable coefficients of Eq. \ref{eq:spmparaVarCoeffs} easily provide models that describe the Higgs mass as well as the neutralino as dark matter candidate. Further, these models escaped LHC measurements, as squarks and gluinos are too heavy ($>$ 1.5 TeV for gluinos, $>$ 2.1 TeV for squarks).

\section{Conclusion}
We investigated supersymmetric models with non-universality in the gaugino sector. This class of models was first introduced by Younkin and Martin \cite{bib:JEYounkinSPMartin}, who investigated a mixing of $SU(5)$'s singlet representation with the \textbf{24} representation. We extended this mixture to the more general case of all representations appearing in the symmetric product of \textbf{24} $\otimes$ \textbf{24}. We focused on the phenomenological implications with respect to the relic density, to recent experimental results from selected direct and indirect detection measurements and to a possible Higgs boson with a mass of $122 < m_{h} < 128$ GeV.\\
\indent The probable detection of the Higgs boson last year puts strongest constraints on the parameter space investigated by Younkin et al. Extending the parameter space by the four mixing angles $\theta_{1}$, $\theta_{24}$, $\theta_{75}$ and $\theta_{200}$ (in contrast to one single mixing angle $\theta$ in \cite{bib:JEYounkinSPMartin}) extends the phenomenological implications of models with non-universal gaugino masses, i.e. provides a solution to this ''Higgs'' problem. We found a factor of $\sim$ 9 more models provide a candidate model with a Higgs boson mass that is consistent with measurements within experimental and theoretical uncertainties. These regions are not constrained to a certain range of angles $\theta_{i}$ but cover the complete simulated range of $\theta_{i}$. Furthermore, models with $\Omega h^{2} < 0.13$ highly coincide with models where $122 < m_{h} < 128$ GeV is respected. Gluino and squark masses are sufficiently high to escape LHC experiments from detection, such that this kind of models are still viable models that can explain a Higgs boson with a mass of 125 GeV/$c^{2}$ and provide the neutralino as a dark matter candidate, given the possibility for dark matter to be produced non-thermally.\\
\indent We performed a detailed study on the dark matter relic density and regions that are excluded by direct and selected indirect detection experiments as well as Higgs boson mass constraints. For simplicity and the sake of clarity we unified the mixing angles $\theta_{1} = \theta_{24} = \theta_{75} = \theta_{200} \equiv \theta$.\\
\indent We found that the parameter space of the considered model can be classified into four regions with respect to the neutralinos composition. These are a pure wino region ($\theta/\pi \lesssim -0.06$ and $\theta/\pi \gtrsim 0.58$ for $\mathrm{tan}\beta =$ 10, $\theta/\pi \lesssim -0.04$ for $\mathrm{tan}\beta =$ 45), and second a pure bino region for $-0.05 \lesssim \theta/\pi \lesssim 0.03$ for both $\mathrm{tan}\beta =$ 10 and 45. In the third region the neutralino is a pure higgsino ($0.06 \lesssim \theta/\pi \lesssim 0.52$ for $\mathrm{tan}\beta =$ 10 and $\theta/\pi \gtrsim 0.06$ for $\mathrm{tan}\beta =$ 45). The last region is characterized by a neutralino that is a wino/higgsino mixture. It can be found for $0.02 \lesssim \theta/\pi \lesssim 0.06$ and $0.52 \lesssim \theta/\pi \lesssim 0.56$ ($\mathrm{tan}\beta =$ 10) and for $0.02 \lesssim \theta/\pi \lesssim 0.06$ ($\mathrm{tan}\beta =$ 45).\\
\indent The relic density in the pure bino region is higher than the upper constraint on $\Omega h^{2}$ ($\Omega h^{2} < 0.13$), analogue to the bulk region of the cMSSM. Concerning the other regions $\Omega h^{2}$ drops rapidly below the lower bound deduced from WMAP data. We showed, that relaxing the WMAP constraint to lower values, allowing the possibility for dark matter to be produced non-thermally, provides significantly more models that satisfy the relaxed relic density requirement compared to the model parameterization introduced by Younkin et al.\\
\indent To obtain a mass of the Higgs boson within $122 < m_{h} < 128$ GeV, $m_{0}$ must be at least 4 TeV and $\theta/\pi$ must be smaller than zero, or bigger than 0.54 for $\mathrm{tan}\beta =$ 10. For $\mathrm{tan}\beta =$ 45, $m_{0}$ must exceed $\sim$ 3.4 TeV and $\theta/\pi$ must be lower than $0.02$. Nevertheless, more models provide a Higgs boson ($\sim$ factor of two for $\mathrm{tan}\beta =$ 10, and $\sim$ factor of 9 for $\mathrm{tan}\beta =$ 45), that satisfies $122 < m_{h} < 128$ GeV compared to the parameterization introduced by \cite{bib:JEYounkinSPMartin}. Almost all of these models ($\mathcal{O}$(95\%)) agree with $\Omega h^{2} < 0.13$.\\
\indent Currently, the best exclusion limit for non-universal models investigated in this work are given by the IceCube and XENON collaborations. IceCube can exclude 8\% and 18.5\% of all models in the channel $W^{+}W^{-}$ for $\mathrm{tan}\beta =$ 10 and $\mathrm{tan}\beta =$ 45, respectively. This means that approximately 20\% (44\%) of models that agree with a Higgs of $122 < m_{h} < 128$ GeV and $\Omega h^{2}$ are excluded for $\mathrm{tan}\beta =$ 10 (45).\\
\indent The current XENON 100 direct detection experiment with a life time of 225 days excludes $\sim$ 11\% of all models for $\mathrm{tan}\beta =$ 10 and $\sim$ 18\% for $\mathrm{tan}\beta =$ 45. This corresponds to approximately 21\% of excluded models with $122 < m_{h} < 128$ GeV and $\Omega h^{2}$ ($\mathrm{tan}\beta =$ 10) and $\sim$ 30\% for $\mathrm{tan}\beta =$ 45. The future XENON 1t, will be even more restrictive. It can exclude $\sim$ 83\% and $\sim$ 76\% of all models, for $\mathrm{tan}\beta =$ 10 and 45,respectively, which means that $\sim$ 78\% and $\sim$ 81\% of models consistent with Higgs and $\Omega h^{2}$ can be tested.\\
\indent For the parameter regions tested in this paper we find that a SU(5) singlet is not able to describe a supersymmetric scenario which is in agreement with the dark matter relic density and the observed Higgs mass. Instead, a mixing of other representations into the singlet allows for models consistent with observations. A linear combination including all non-singlet representations of SU(5) that appear in the symmetric product of \textbf{24} $\otimes$ \textbf{24} cannot be excluded by current measurements from direct and the considered indirect detection methods. With the newly detected particle at the LHC being the Higgs boson reduces the possible parameter space, but neither existing data from dark matter search nor predicted sensitivities, e.g. XENON 1t, can ultimately exclude the model investigated in this work.\\
\indent Last but not least we did a preliminary analysis on the gaugino mass ratios. We allowed a variable mass ratio between the gaugino mass parameters $M_{1}$, $M_{2}$ and $M_{3}$. Therefore, we varied the coefficients (a,b,c) that determine the ratios of $M_{i}$ at a given mixing angle $\theta$ at the GUT scale. We found that a Higgs boson with a mass of $122 < m_{h} < 128$ GeV is easily achieved, even at lower values of $m_{0} = 3$ TeV compared to $m_{0} \gtrsim 4$ TeV with respect to the other linear combinations. Furthermore, it is possible to require more stringent constraints on the relic density, e.g. $0.09 < \Omega h^{2} < 0.13$. A reasonable amount of models with these constraints remain and provides candidate models for dark matter produced thermally while simultaneously satisfying constraints from the Higgs boson.

\acknowledgments{We would like to thank S.P. Martin for various fruitful discussions. This work was partially funded by the German Ministry of Education and Research (BMBF) contract number 05A11WEA.}


\begin{thebibliography}{999}
  \bibitem{bib:CMSHiggs} The CMS Collaboration, {\it Observation of a new boson with mass near 125 GeV in pp collisions at sqrt(s) = 7 and 8 TeV}, 2013 arXiv:1303.4571v1 [hep-ex]
 \bibitem{bib:ATLASHiggs} The ATLAS Collaboration, {\it Observation of a New Particle in the Search for the Standard Model Higgs Boson with the ATLAS Detector at the LHC}; 2012 Phys. Lett. B716 (2012) 1-29
 \bibitem{bib:JEYounkinSPMartin} Younkin J E and Martin S P, {\it Non-universal gaugino masses, the supersymmetric little hierarchy problem, and dark matter}, 2012 arxiv:1201.2989 [hep-ph]
 \bibitem{bib:CremmerEtAl} Cremmer E, Julia B, Ferrara S, Girardello L and van Proeyen A, {\it Coupling Supersymmetric Yang-Mills Theories to Supergravity}, 1982 Phys. Lett. B116 (1982) 231
 \bibitem{bib:HuituEtAl} Huitu K, Kawamura Y, Kobayashi T and Puolam\"aki K, {\it Phenomenological Constraints on SUSY SU(5) GUTs with Non-universal Gaugino Masses}, 2000 Phys. Rev. D61 (2000) 035001 [arXiv:hep-ph/9903528]
 \bibitem{bib:EllisEtAl} Ellis J, Enqvist K, Nanopoulos D V and Tamavakis K, {\it Gaugino Masses And Grand Unification}, 1985 Phys. Lett. B155 (1985) 351
 \bibitem{bib:AmundsonEtAl} Amundson J et al, {\it Report of the supersymmetry theory subgroup}, 1996 arxiv:hep-ph/9609374
 \bibitem{bib:DjouadiKneurMoultaka} Djouadi A, Kneur J L and Moultake G, {\it SuSpect: a Fortran Code for the Supersymmetric and Higgs Particle Spectrum in the MSSM}, 2007 Comput. Phys. Commun 176:426-455 [hep-ph/0211331]
 \bibitem{bib:GondoloEdsjoBaltz} Gondolo P, Edsj\"o J, Ullio P, Bergstr\"om L, Schelke M and Baltz E A, {\it DarkSUSY: Computing Supersymmetric Dark Matter Properties Numerically}, 2004 JCAP 07 008 [astro-ph/0406204], Gondolo P, Edsj\"o J, Ullio P, Bergstr\"om L, Schelke M, Baltz E A, Bringmann T and Duda G, http://www.darksusy.org.
 \bibitem{bib:PdG}Nakamura K et al. [Particle Data Group Collaboration], 2010 and 2011 J. Phys. G 37, 075021 (2010) and 2011 partial update for the 2012 edition
 \bibitem{bib:TMoroiLRandall} Moroi T and Randall L, {\it Wino Cold Dark Matter from Anomaly-Mediated SUSY Breaking}, 2000 Nucl. Phys. B570:455-472 [arXiv:hep-ph/9906527]
 \bibitem{bib:DChungEKolbARiotto} Chung D J H, Kolb E W and Riotto A, {\it Production of massive particles during reheating}, 1999 Phys. Rev. D60 063504 [arXiv:hep-ph/9809453]
 \bibitem{bib:RJeannerotEtAl} Jeannerot R, Zhang X and Brandenberger R, {\it Non-thermal Production of Neutralino Cold Dark Matter from Cosmic String Decays}, 1999 JHEP12(1999)003 [arXiv:hep-ph/9901357]
 \bibitem{bib:ATLASSquarks} The ATLAS Collaboration, {\it Search for squarks and gluinos with the ATLAS detector using final states with jets and missing transverse momentum at $\sqrt{s} =$ 8 TeV}, 2012 ATLAS-CONF-2012-109, retrieved from http://cds.cern.ch/record/1472710/files/ATLAS-CONF-2012-109.pdf
 \bibitem{bib:FrereErAl} Frere J M, Jones D R T and Raby S, {\it Fermion masses and induction of the weak scale by supergravity}, 1983 Nucl. Phys B222 (1983) 11
 \bibitem{bib:AlvarezEtAl} Alvarez-Gaume L, Polchinski J and Wise M B, {\it Minimal low-energy supergravity} 1983 Nucl. Phys. B221 (1983) 495
 \bibitem{bib:DerendingerEtAl} Derendinger J P and Savoy C A, {\it Quantum effects and SU(2)$\times$U(1) breaking in supergravity gauge theories}, 1984 Nucl. Phys. B237 (1984) 307
 \bibitem{bib:KounnasEtAl} Kounnas C, Lahanas A B, Nanopoulos D V and Quiros M, {\it Low-energy behaviour of realistic locally-supersymmetric grand unified theories}, 1984 Nucl. Phys. B236 (1984) 438
 \bibitem{bib:BAllanachADjouadiEtAl} Allanach B C, Djouadi A, Kneur J L, Porod W and Slavich P, {\it Precise determination of the neutral Higgs boson mass in the MSSM}, 2004 JHEP0409:044 [arXiv:hep-ph/0406166]
 \bibitem{bib:ZCharifVBertinPGay} The ANTARES Collaboration, {\it First Search for Dark Matter Annihilation in the Sun Using the ANTARES Neutrino Telescope}, 2013 arXiv:1302.6516 [astro-ph.HE]
 \bibitem{bib:IceCube} The IceCube Collaboration, {\it The IceCube Neutrino Observatory IV: Searches for Dark Matter and Exotic Particles}, 2011 arxiv:1111.2738 [astro-ph.HE]
 \bibitem{bib:CDMSII} Ahmed Z et al. (The CDMS Collaboration), {\it Dark Matter Search Results from the CDMS II Experiment}, 2010 Science 327 (2010) 1619-1621
 \bibitem{bib:Edelweiss} Armengaud E et al. (EDELWEISS Collaboration), {\it Final results of the EDELWEISSII-WIMP search using a 4-kg array cryogenic germanium detectors with interleaved electrodes}, 2011 Physics Letters B 702 (2011) 329-335
 \bibitem{bib:XENON100} April E et al. (The XENON100 Collaboration), {\it Dark Matter Results from 225 Live Days of XENON100 Data}, 2012 arxiv:1207.5988 [astro-ph.CO]
 \bibitem{bib:XENON1T} April E et al. (The XENON1T Collaboration); {\it The XENON 1T Dark Matter Search Experiment}, 2012 arxiv:1206.6288 [astro-ph.IM]
 \bibitem{bib:JDWells} Wells J D, {\it PeV-Scale Supersymmetry}, 2005 Phys. Rev. D71 (2005) 015013
\end{thebibliography}
\end{document}